\documentclass[fleqn,usenatbib]{mnras}

\usepackage{newtxtext,newtxmath}

\usepackage[T1]{fontenc}

\DeclareRobustCommand{\VAN}[3]{#2}
\let\VANthebibliography\thebibliography
\def\thebibliography{\DeclareRobustCommand{\VAN}[3]{##3}\VANthebibliography}

\usepackage{graphicx}	
\usepackage{amsmath}	

\makeatletter 
  \patchcmd{\NAT@citex}
    {\@citea\NAT@hyper@{%
      \NAT@nmfmt{\NAT@nm}%
      \hyper@natlinkbreak{\NAT@aysep\NAT@spacechar}{\@citeb\@extra@b@citeb}%
      \NAT@date}}
    {\@citea\NAT@nmfmt{\NAT@nm}%
    \NAT@aysep\NAT@spacechar\NAT@hyper@{\NAT@date}}{}{}

  \patchcmd{\NAT@citex}
    {\@citea\NAT@hyper@{%
      \NAT@nmfmt{\NAT@nm}%
      \hyper@natlinkbreak{\NAT@spacechar\NAT@@open\if*#1*\else#1\NAT@spacechar\fi}%
        {\@citeb\@extra@b@citeb}%
      \NAT@date}}
    {\@citea\NAT@nmfmt{\NAT@nm}%
    \NAT@spacechar\NAT@@open\if*#1*\else#1\NAT@spacechar\fi\NAT@hyper@{\NAT@date}}
    {}{}
\makeatother

\usepackage{ae,aecompl}
\usepackage{times}
\usepackage{enumitem}
\usepackage{times}
\usepackage{textcomp}
\usepackage{verbatim}
\usepackage{esdiff}
\usepackage{xspace}

\newcommand{\msun}{{\,\rm M_\odot}}

\newcommand{\Gyr}{\,{\rm Gyr}}

\newcommand{\Myr}{\,{\rm Myr}}
\newcommand{\pc}{\,{\rm pc}}
\newcommand{\kpc}{\,{\rm kpc}}

\newcommand{\Mpc}{\,{\rm Mpc}}

\newcommand{\mmag}{\,{\rm mag}}
\newcommand{\kev}{\,{\rm keV}}

\newcommand\thesan{\mbox{\textsc{thesan}}\xspace}
\newcommand\thesanone{\mbox{\textsc{thesan-1}}\xspace}
\newcommand\thesanhr{\mbox{\textsc{thesan-hr}}\xspace}
\newcommand\thesanhrlarge{\mbox{\textsc{thesan-hr-large}}\xspace}
\newcommand\thesanhrwdm{\mbox{\textsc{thesan-hr-wdm}}\xspace}
\newcommand\thesanhrsdao{\mbox{\textsc{thesan-hr-sdao}}\xspace}
\newcommand\thesanhrfdm{\mbox{\textsc{thesan-hr-fdm}}\xspace}
\newcommand\thesanhruvb{\mbox{\textsc{thesan-hr-uvb}}\xspace}
\newcommand\thesanhruvbwdm{\mbox{\textsc{thesan-hr-uvb-wdm}}\xspace}
\newcommand\thesanhruvbsdao{\mbox{\textsc{thesan-hr-uvb-sdao}}\xspace}
\newcommand\thesanhruvbfdm{\mbox{\textsc{thesan-hr-uvb-fdm}}\xspace}

\def\jcap{J. Cosmol.  Astropart. Phys.}
\def\aap{A\&A}
\def\apj{ApJ}

\def\apjl{ApJ}
\def\mnras{MNRAS}
\def\araa{ARA\&A}
\def\aj{AJ}

\def\na{New Astronomy}

\def\physrep{Phys. Rep.}
\def\nat{Nature}

\def\apjs{ApJS}
\def\prd{Phys. Rev. D}

\title[Alternative DM in \thesanhr]{\thesanhr: Galaxies in the Epoch of Reionization in warm dark matter, fuzzy dark matter and interacting dark matter}

\author[Shen et al.]{\parbox{17.5cm}{
Xuejian Shen,$^{1}$\thanks{E-mail: \href{mailto:xshen@caltech.edu}{xshen@caltech.edu}}
Josh Borrow,$^{2}$
Mark Vogelsberger,$^{2,3}$
Enrico Garaldi,$^{4}$
Aaron Smith,$^{5}$
Rahul Kannan,$^{6}$
Sandro Tacchella,$^{7,8}$
Jes\'{u}s Zavala,$^{9}$
Lars Hernquist,$^{5}$
Jessica Y.-C. Yeh$^{2}$
and
Chunyuan Zheng$^{10}$
\\
}\vspace{0.3cm}\\
$^{1}$ TAPIR, California Institute of Technology, Pasadena, CA 91125, USA \\
$^{2}$ Department of Physics \& Kavli Institute for Astrophysics and Space Research, Massachusetts Institute of Technology, Cambridge, MA 02139, USA \\
$^{3}$ The NSF AI Institute for Artificial Intelligence and Fundamental Interactions, Massachusetts Institute of Technology, Cambridge, MA 02139, USA \\
$^{4}$ Max-Planck Institute for Astrophysics, Karl-Schwarzschild-Str. 1, D-85741 Garching, Germany \\
$^{5}$ Center for Astrophysics | Harvard \& Smithsonian, 60 Garden Street, Cambridge, MA 02138, USA \\
$^{6}$ Department of Physics and Astronomy, York University, 4700 Keele Street, Toronto, ON M3J 1P3, Canada \\
$^{7}$ Kavli Institute for Cosmology, University of Cambridge, Madingley Road, Cambridge, CB3 0HA, UK \\
$^{8}$ Cavendish Laboratory, University of Cambridge, 19 JJ Thomson Avenue, Cambridge, CB3 0HE, UK \\
$^{9}$ Center for Astrophysics and Cosmology, Science Institute, University of Iceland, Dunhagi 5, 107 Reykjavik, Iceland \\
$^{10}$ University of California Berkeley, Berkeley, CA 94720, USA \\
}

\date{Accepted XXX. Received YYY; in original form ZZZ}

\pubyear{2023}

\begin{document}
\label{firstpage}
\pagerange{\pageref{firstpage}--\pageref{lastpage}}
\maketitle

\begin{abstract}
Using high-resolution cosmological radiation-hydrodynamic (RHD) simulations (\thesanhr), we explore the impact of alternative dark matter (altDM) models on galaxies during the Epoch of Reionization. The simulations adopt the IllustrisTNG galaxy formation model. We focus on altDM models that exhibit small-scale suppression of the matter power spectrum, namely warm dark matter (WDM), fuzzy dark matter (FDM), and interacting dark matter (IDM) with strong dark acoustic oscillations (sDAO). In altDM scenarios, both the halo mass functions and the UV luminosity functions at $z\gtrsim 6$ are suppressed at the low-mass/faint end, leading to delayed global star formation and reionization histories. However, strong non-linear effects enable altDM models to ``catch up'' with cold dark matter (CDM) in terms of star formation and reionization. The specific star formation rates are enhanced in halos below the half-power mass in altDM models. This enhancement coincides with increased gas abundance, reduced gas depletion times, more compact galaxy sizes, and steeper metallicity gradients at the outskirts of the galaxies. These changes in galaxy properties can help disentangle altDM signatures from a range of astrophysical uncertainties. Meanwhile, it is the first time that altDM models have been studied in RHD simulations of galaxy formation. We uncover significant systematic uncertainties in reionization assumptions on the faint-end luminosity function. This underscores the necessity of accurately modeling the small-scale morphology of reionization in making predictions for the low-mass galaxy population. Upcoming {\itshape James Webb Space Telescope} (JWST) imaging surveys of deep, lensed fields hold potential for uncovering the faint, low-mass galaxy population, which could provide constraints on altDM models.
\end{abstract}




\section{Introduction}

The nature of dark matter (DM) is one of the most outstanding questions in modern cosmology~\citep[see][for a review]{Bertone2018_dmreview}. In the standard cosmological model, DM is assumed to be cold after its decoupling from the primordial plasma and is effectively collisionless (known as CDM). This model is successful at describing cosmic structure formation on large scales ($\gtrsim 1 \Mpc$; e.g. \citealt{Blumenthal1984,Davis1985}) and acts as the foundation for the theory of galaxy formation~\citep[e.g.][]{White1978,Dekel1986,Kauffmann1993,Springel2005,Behroozi2013,Vogelsberger2014,Schaye2015}. Many beyond-standard-model theories do predict DM candidates with such properties (see \citealt{Bertone2005} for a review), for example, the weakly-interacting massive particles (WIMPs) motivated by the hierarchy problem as well as the light axion(-like) particles (ALPs) motivated by the strong CP problem and string theories. The success of CDM on astrophysical scales and the natural production mechanisms of these candidates have motivated decades of experimental searches~\citep[e.g.][]{Ahmed2009_FIVE-TOWER,CDMS2010,Aprile2012_XENON100,Akerib2014_LUX,Akerib2017_LUX,Aprile2018_XENON1T}. However, as the parameter space and model configuration of these CDM candidates become increasingly constrained given the null results in direct detection~\citep[e.g.][]{Aprile2018_XENON1T, Bertone2018_wimpalt}, there are strong motivations to explore DM candidates beyond the CDM paradigm. Meanwhile, on small scales (Local dwarf galaxies), several discrepancies between CDM predictions and astrophysical observations have come to light \citep[see][for a review]{Bullock2017}, including but not limited to the {\itshape missing satellites} problem~\citep[e.g.][]{Klypin1999, Moore1999}, the {\itshape too-big-to-fail} problem~\citep[TBTF;][]{MBK2011,MBK2012,Tollerud2014}, the {\itshape core-cusp} problem~\citep[e.g.][]{Flores1994,Moore1994,deBlok2001,KDN2006,Oh2015}, and recently the diversity problem~\citep[e.g.][]{Oman2015,Kaplinghat2019,Zavala2019}. DM models alternative to the classical CDM has been proposed as the solution to these anomalies in the Local Universe. However, despite the application to solve the local puzzles, a category of alternative DM (altDM~\footnote{Not to be confused with physics models alternative to DM.} hereafter) models can modify the initial linear matter power spectrum at small, poorly-constrained scales ($k\gtrsim 10\,h/{\rm Mpc}$) which determines the initial conditions of non-linear structure formation. Such modifications can affect the assembly of haloes/galaxies in the early Universe as well as cosmic reionization. In the following, we will introduce three DM models that fall into this category. 

\begin{itemize}
    \item Warm dark matter (WDM) is a class of DM candidates with large free-streaming velocities. The free-streaming of WDM can suppress the density fluctuations on $\Mpc$ scales and below. WDM can be produced through the freeze-out mechanism after reaching thermal equilibrium with the initial plasma \citep[e.g. gravitino;][]{Steffen2006}. In addition to the thermal relic WDM, the sterile neutrino is another well-motivated candidate, which can be non-thermally produced through scattering processes due to their mixing with active neutrinos with the Dodelson--Widrow mechanism \citep{Dodelson1994}, through resonant production \citep{Shi1999}, or through coupling with other fields \citep{Kusenko2006, Shaposhnikov2006}. WDM has been empirically motivated as a solution to some of the above-mentioned small-scale astrophysical challenges \citep[e.g.][]{Bode2001,Lovell2012,Lovell2014,Polisensky2014,Lovell2017}. The latest constraints on WDM (reported at the $2\sigma$ level) are e.g. (equivalent) thermal relic mass $m_{\rm WDM} > 5.3\kev$ ($3.5\kev$, if allowing sudden temperature changes) from Lyman-$\alpha$ forest observations \citep{Irsic2017}, $>5.2\kev$ from substructures of strong gravitational lenses \citep{Gilman2020}, $>6.5\kev$ from the ultra-faint dwarf (UFD) satellites of the Milky Way \citep{Nadler2021} ($\gtrsim 2\kev$ in \citealt{Newton2021}). In this work, we choose to study the benchmark model with $m_{\rm WDM}=3\kev$. It lies at the edge of the existing constraints bearing in mind the inevitable model-dependence and systematical uncertainties of all reported observational constraints. This value gives prominent signals at the halo mass scale that can be well-resolved by our simulations (see Section~\ref{sec:sim} for more discussions) and is of interest in explaining the $3.55\kev$ X-ray emission from the Galatic center as the decay product of sterile neutrinos \citep[e.g.][]{Bulbul2014,Boyarsky2014,Schneider2016,Adhikari2017}. 

    \item Dark Acoustic Oscillations (DAOs) can appear for a generic class of interacting dark matter (IDM) models~\footnote{Not all IDM models feature DAOs. We are studying an empirical benchmark model with strong DAOs in this work.}. The acoustic oscillations can be caused by interactions between DM and Standard Model particles in the early Universe \citep[e.g.][]{Boehm2002, Sigurdson2004, Boehm2005, Boehm2014, Schewtschenko2016, Boddy2018, Gluscevic2018} and has been considered for thermal WIMP particles \citep[e.g.][]{Loeb2005,Bertschinger2006,Bringmann2009}. It was then also generalized to hidden-sector DM models with e.g. DM–dark radiation interactions \citep[e.g.][]{vandenAarssen2012, Buckley2014, Cyr-Racine2014}, which has been included in the Effective Theory of Structure Formation (ETHOS) framework \citep{Cyr-Racine2016, Vogelsberger2016}. DAOs take place on small scales and are accompanied by collisional damping, which can suppress the abundance of small-scale structures as in WDM but with a different physical mechanism. As a continuation of the ETHOS work, \citet{Bohr2020} proposed effective parameters of DAO models that fully characterize structure formation in the non-linear regime at high redshifts ($z\geq 5$) in a simple and clear way. We pick the strong DAO (sDAO) model studied in \citet{Bose2019} and \citet{Bohr2021} with the relative strength of the first DAO peak set to unity. This represents a model with strong small-scale residual fluctuations in addition to the major suppression feature at a slightly larger scale.
    
    \item Fuzzy dark matter (FDM) consists of ultralight axion-like particles (ALPs) with a typical mass of $m_{\rm a}\lesssim 10^{-20}\,{\rm eV}$ with strong particle physics theory motivations \citep[e.g.][]{Preskill1983,Svrcek2006,Sikivie2009,Arvanitaki2010}. These bosonic particles with astrophysical de Broglie wavelengths generate effective quantum pressure on small scales and can suppress structural formation. Cosmological and astrophysical consequences of FDM have been comprehensively reviewed in \citet{Hu2000}, \citet{Hlozek2015}, \citet{Marsh2016} and \citet{Hui2017}. The latest constraints on FDM (reported at the $2\sigma$ level) are e.g. ALP mass $m_{\rm a}> 20 \times 10^{-22}\,{\rm eV}$ from Lyman-$\alpha$ forest observations \citep{Irsic2017b}, $\gtrsim 8\times 10^{-22}\,{\rm eV}$ from galaxy ultraviolet (UV) luminosity function \citep{Menci2017,Ni2019}, $>29\times 10^{-22}\,{\rm eV}$ from the UFD satellites of the Milky Way \citep{Nadler2021}, $\gtrsim 10^{-19}\,{\rm eV}$ from core-oscillations of the UFD Eridanus {\small II} \citep{Marsh2019}, and $> 1-3\times 10^{-19}\,{\rm eV}$ from stellar kinematics of the Segue 1 and Segue 2 UFDs \citep{Hayashi2021,Dalal2022}. We pick $m_{\rm a} = 2\times 10^{-21}\,{\rm eV}$ for our simulations so that it provides a similar damping wavenumber to the benchmark $3\kev$ WDM model studied.
\end{itemize}

Structure formation at high redshift serves as a promising complementary channel to constrain the particle nature of DM. The population of faint galaxies during the Epoch of Reionization (EoR) will be sensitive to the altDM physics that suppresses the small-scale power spectrum. The population of high-redshift galaxies has been studied in the context of WDM \citep[e.g.][]{Schultz2014, Bose2016, Dayal2017, Lopez2017, Lovell2018, Menci2018, Bozek2019}, FDM \citep[e.g.][]{Bozek2015, Menci2017, Ni2019, Mocz2020} and DAO models \citep{Lovell2018, Bohr2021, Kurmus2022}. The methods adopted include DM-only simulations with abundance matching \citep[e.g.][]{Schultz2014,Corasaniti2017}, semi-analytical models of galaxy formation \citep[e.g.][]{Bose2016,Dayal2017,Khimey2021} as well as cosmological hydrodynamic simulations \citep[e.g.][]{Lovell2018,Ni2019}. 

The suppression of the rest-frame UV luminosity function was identified at faint magnitudes $M_{\rm UV} \gtrsim -16$ in these models. The UV luminosity functions probed by deep {\itshape Hubble Space Telescope} (HST) programs \citep[e.g.][]{Livermore2017,Atek2018,Ishigaki2018,Bouwens2022} have been used to constrain altDM models that suppress small-scale structure formation. In the near future, deep imaging surveys of strongly lensed, inherently faint galaxies concluded with the {\itshape James Webb Space Telescope} (JWST) will push the detection limit by at least three magnitudes and extend the redshift frontier of galaxy searches with its infrared frequency coverage, providing compelling tests on the nature of DM. For example, as part of the JWST Early Release Observations (ERO; \citealt{Pontoppidan2022}), the galaxy cluster SMACS0723 was imaged using NIRCam, yielding the discovery of galaxies out to $z\sim 16$ \citep[e.g.][]{Atek2023}. Another Early Release Science (ERS) program (GLASS-JWST; \citealt{Treu2022}) pointed to one foreground Hubble Frontier Field (HFF) cluster, Abell 2744. Forthcoming observational programs will extend the measurements of faint-end luminosity functions in these lensed fields. Besides galaxy abundances, many theoretical studies \citep[e.g.][]{Bose2016,Bose2017,Corasaniti2017,Dayal2017,Lovell2018,Lovell2019,Ni2019} found higher occupation fraction of luminous galaxies, higher UV luminosities at a given halo mass in altDM models than in the CDM case. Despite the initial delay in the assembly of DM haloes, the stellar content of galaxies undergoes faster build-up in altDM models, which closes the gap in the global star formation and reionization history between models at lower redshift. The strong, rapid starbursts in low-mass galaxies in these models could be another interesting feature for observations of high-redshift galaxies as well as the Local Group UFDs (e.g. \citet{Lovell2019} and \citet{Bozek2019} found that the simulated dwarfs have delayed but more rapid and diverse star formation histories in WDM). 

In addition to the galaxy properties, the neutral gas distribution in the intergalactic medium (IGM) will be sensitive to these altDM physics as well. The mass distribution of baryonic gas is a tracer of the DM density field. The phase of the gas has additional dependencies on the intergalactic ionizing radiation from star formation in galaxies. The neutral phase of the gas can be mapped by the Lyman-$\alpha$ forest observations at $z \lesssim 5-6$ \citep[e.g.][]{Cen1994, Hernquist1996, Seljak2005, Viel2005, Viel2013, Delubac2015,Yang2020,Bosman2022} and the intensity mapping of the 21 cm spin-flip transition of the hydrogen atom \citep{Furlanetto2006, Mellema2006, Parsons2010, Pritchard2012}. The global signal of the ionized phase can be constrained by the optical depth of the Cosmic Microwave Background (CMB). The impact of altDM models on the IGM properties and various detection signals has been studied in e.g. \citet{Viel2013, Das2018, Lovell2018, Bose2019, Munoz2021, Munoz2022}. In the near future, a more complete picture of the neutral gas distribution in the Universe will be provided by current and upcoming radio instruments, including the Low-Frequency Array (LOFAR; \citealt{vanHaarlem2013-LOFAR}), Square Kilometer Array (SKA; \citealt{Dewdney2009-SKA}), and Hydrogen Epoch of Reionization Array (HERA; \citealt{DeBoer2017-HERA}). These instruments will constrain the timing and morphology of reionization, the properties of the first galaxies, the evolution of large-scale structure, and the early sources of heating throughout the EoR and cosmic dawn. For example, as highlighted in \citet{Munoz2021}, the 21 cm observation can constrain the small-scale fluctuations at $k \lesssim 300\,h/\Mpc$, which has never been reached by any other constraints.

All the EoR constraints involve non-negligible astrophysical uncertainties. This includes the escape fraction of ionizing photons, the star formation efficiency regulated by feedback from star formation as well as the physics that sets the atomic cooling limit for galaxy formation. For example, radiative and supernovae feedback from both Population {\small III} (Pop{\small III}) and metal-enriched Pop{\small II} stars can regulate or even prevent star formation in low-mass galaxies and flatten the faint-end UV luminosity function \citep[e.g.][]{Jaacks2013,Wise2014,Oshea2015,Xu2016,Ocvirk2016,Dayal2018}. This can contaminate the signal of altDM physics. The delay in reionization in altDM models can also be compensated by a steeper redshift evolution of the ionizing photon escape fraction \citep[e.g.][]{Dayal2017}. Therefore, it is essential to have theoretical forecasts based on robust galaxy formation and reionization models that have been tested against a wide range of observational constraints. The \thesan project \citep{Kannan2022,Garaldi2022,Smith2022} is a suite of radiation-magneto-hydrodynamic simulations featuring an efficient radiation hydrodynamics solver \citep{Kannan2019} that precisely captures the interaction between ionizing photons and gas. It is coupled to the well-tested IllustrisTNG galaxy formation model~\citep{Pillepich2018, Nelson2019}. \thesan simulations have been used to make a wide range of predictions for the Universe in the EoR \citep[e.g.][]{Smith2022,Garaldi2022,Kannan2022,Kannan2022b,Kannan2022c,Borrow2022,Xu2022,Yeh2023}. In this work, we use the small-volume, high-resolution variant of the \thesan simulations (namely the \thesanhr project introduced in \citealt{Borrow2022}) to study galaxy formation in the EoR in altDM models with self-consistent modeling of the reionization process at small scales.

This paper is organized as follows. In Section~\ref{sec:sim}, we introduce the simulations and the altDM models considered. A visual overview of the DM density field and IGM properties is given in Section~\ref{sec:visual}. In Sections~\ref{sec:hmf-smf} and~\ref{sec:uvlf}, we present predictions of galaxy abundance, including the halo/stellar mass function and rest-frame UV luminosity functions. In Section~\ref{sec:sfrd}, we take a deeper dive into galaxy properties in different DM models, discussing scaling relations and spatially resolved properties of galaxies. In Section~\ref{sec:discuss}, we discuss potential ways to disentangle altDM signatures with various uncertainties in astrophysical processes. The conclusions of the paper are given in Section~\ref{sec:conclusion}. Throughout this paper, we employ the \citet{Planck2016} cosmology with $h = 0.6774$, $\Omega_{\rm 0} = 0.3089$, and $\Omega_b = 0.0486$, $\sigma_{8} = 0.8159$, $n_{\rm s} = 0.9667$. 

\begin{table*}
    \addtolength{\tabcolsep}{2.4pt}
    \centering
    \begin{tabular}{lccccccccc}
        \hline
        Simulation & $L_{\rm box}$ & $N_{\rm part}$ & $m_{\rm dm}$ & $m_{\rm baryon}$ & $\epsilon_{\rm dm,\,\ast}$ & $h_{\rm b}$ & Reionization & DM & $k_{\rm 1/2}$ \\
        Name & $[{\rm cMpc}]$ &  & $[\msun]$ & $[\msun]$ & $[{\rm ckpc}]$ & $[{\rm ckpc}]$ & Model & Model & [$h$/Mpc] \\
        \hline
        \hline
        \thesanone & 95.5 & $2\times 2100^{3}$ & $3.12\times10^{6}$ & $5.82\times10^{5}$ & $2.2$ & $2.2$ & \thesan (RT) & CDM & --\\
        \thesanhrlarge & 11.8 & $2\times 512^{3}$ & $4.82\times 10^{5}$ & $9.04\times 10^{4}$ & $0.85$ & $0.85$ & \thesan (RT) & CDM & --\\
        \thesanhr & 5.9 & $2\times 256^{3}$ & $4.82\times 10^{5}$ & $9.04\times 10^{4}$ & $0.85$ & $0.85$ & \thesan (RT) & CDM & --\\
        \thesanhrwdm & 5.9 & $2\times 256^{3}$ & $4.82\times 10^{5}$ & $9.04\times 10^{4}$ & $0.85$ & $0.85$ & \thesan (RT) & WDM & 22\\
        \thesanhrsdao & 5.9 & $2\times 256^{3}$ & $4.82\times 10^{5}$ & $9.04\times 10^{4}$ & $0.85$ & $0.85$ & \thesan (RT) & sDAO & 16\\
        \thesanhrfdm & 5.9 & $2\times 256^{3}$ & $4.82\times 10^{5}$ & $9.04\times 10^{4}$ & $0.85$ & $0.85$ & \thesan (RT) & FDM & 25\\
        \hline
        \thesanhruvb & 5.9 & $2\times 256^{3}$ & $4.82\times 10^{5}$ & $9.04\times 10^{4}$ & $0.85$ & $0.85$ & Uniform UVB & CDM & --\\ 
        \thesanhruvbwdm & 5.9 & $2\times 256^{3}$ & $4.82\times 10^{5}$ & $9.04\times 10^{4}$ & $0.85$ & $0.85$ & Uniform UVB & WDM & 22\\
        \thesanhruvbsdao & 5.9 & $2\times 256^{3}$ & $4.82\times 10^{5}$ & $9.04\times 10^{4}$ & $0.85$ & $0.85$ & Uniform UVB & sDAO & 16\\
        \thesanhruvbfdm & 5.9 & $2\times 256^{3}$ & $4.82\times 10^{5}$ & $9.04\times 10^{4}$ & $0.85$ & $0.85$ & Uniform UVB & FDM & 25\\
        \hline
    \end{tabular}
    \caption{ \textbf{Simulations of the \thesanhr suite. Each column corresponds to the following information:} \newline \hspace{\textwidth}
    (\textbf{1}) Name of the simulation. \thesanhrlarge is the run with identical physics and numeric resolution as \thesanhr but doubling the box side-length.
    (\textbf{2}) $L_{\rm box}$: Side-length of the periodic simulation box. The unit is comoving Mpc (cMpc). 
    (\textbf{3}) $N_{\rm part}$: Number of particles (cells) in the simulation. In the initial conditions, there are an equal number of DM particles and gas cells. 
    (\textbf{4}) $m_{\rm dm}$: Mass of DM particles, which is conserved over time. 
    (\textbf{5}) $m_{\rm baryon}$: Mass of gas cells in the initial conditions as a reference for the baryonic mass resolution. The gas cells are (de-)refined so that the gas mass in each cell is within a factor of two of this target gas mass. Stellar particles stochastically generated out of gas cells can take arbitrary initial masses and are subject to mass loss via stellar evolution~\citep{Vogelsberger2013}. 
    (\textbf{6}) $\epsilon_{\rm dm,\,\ast}$: The gravitational softening length for the DM and stellar particles. 
    (\textbf{7}) $h_{\rm b}$: The minimum gravitational softening length of gas cells, which are adaptively softened. 
    (\textbf{8}) Reionization model used. The fiducial \thesan model employs on-the-fly radiative transfer of ionizing photons while the uniform UVB model assumes a spatially uniform, time-varying background radiation of ionizing photons.
    (\textbf{9}) DM model employed.
    (\textbf{10}) The characteristic wavenumber of the small-scale suppression on the linear matter power spectrum, defined as $P_{\rm altDM}(k_{\rm 1/2}) = 1/2 \,P_{\rm CDM}(k_{\rm 1/2})$.
    }
    \label{tab:sims}
    \addtolength{\tabcolsep}{-2.4pt}
\end{table*}

\section{Simulations}
\label{sec:sim}

The \thesan project \citep{Kannan2022,Garaldi2022,Smith2022} is a suite of radiation-magneto-hydrodynamic simulations utilizing the moving-mesh hydrodynamics code {\sc AREPO} \citep{Springel2010, Weinberger2020}. Gravity is solved using the hybrid Tree-PM method \citep{Barnes1986}. The hydrodynamics is solved using the quasi-Lagrangian Godunov method \citep{Godunov1959} on an unstructured Voronoi mesh grid \citep[see][for a review]{Vogelsberger2020NatR}. For self-consistent treatment of ionizing radiation, the \thesan project employs the radiative transfer (RT) extension {\sc Arepo-rt} \citep{Kannan2019}, which solves the first two moments of the RT equation assuming the M1 closure relation \citep{Levermore1984}. The simulation includes the sourcing (from stars and AGNs) and propagation of ionizing photons (in three energy bins relevant for hydrogen and helium photoionization between energy intervals of $[13.6, 24.6, 54.4, \infty)\,{\rm eV}$) as well as a non-equilibrium thermochemistry solver to model the coupling of radiation fields to gas. The luminosity and spectral energy density of stars in \thesan as a complex function of age and metallicity are calculated using the Binary Population and Spectral Synthesis models (BPASS v2.2.1; \citealt{Eldridge2017}). The sub-grid escape fraction of stars was set to be $f_{\rm esc} = 0.37$ in the simulations to match the global reionization history of the Universe in CDM. For details of the simulation methods, we refer to \citet{Kannan2019, Kannan2022}. 

In terms of the galaxy formation model, the simulations employ the IllustrisTNG model~\citep{Pillepich2018, Nelson2019}, which is an update of the Illustris model \citep{Vogelsberger2014,Vogelsberger2014b}. The simulations include (1) density-, temperature-, metallicity- and redshift-dependent cooling of metal-enriched gas, (2) a two-phase, effective equation of state model for the interstellar medium (ISM) at the sub-resolution level~\citep{Springel2003}, (3) star formation in dense gas following the empirically defined Kennicutt–Schmidt relation, (4) thermal and mechanical feedback from supernovae and stellar winds, (5) metal enrichment from stellar evolution and supernovae, and (6) SMBH formation, growth, and feedback \citep[in two regimes described in][although SMBH physics have little impact on the low-mass galaxies studied in \thesanhr]{Weinberger2017}. The model has been extensively tested in large-scale simulations and is able to produce realistic galaxies that match a wide range of observations~\citep[e.g.][]{Nelson2018, Springel2018, Genel2018, Pillepich2018b, Naiman2018, Vogelsberger2018, Marinacci2018, Vogelsberger2020, Shen2020}. 

\citet{Borrow2022} introduced a subset of high-resolution small-volume simulations (\thesanhr) using the same numeric setup and physics inputs as the main \thesan suite, aiming to explore the formation and evolution of low-mass galaxies in the early Universe. The mass resolution of DM (gas) is set to $4.82\times 10^{5}\msun$ ($9.04\times 10^{4}\msun$), which is about $6.4$ times better than the flagship \thesanone simulation and allows atomic cooling haloes ($M_{\rm halo} \simeq 10^{7.5-8}\msun$; \citealt{Wise2014}) to be properly resolved ($\gtrsim 100$ DM particles). The gravitational softening lengths of DM and stars are set to $0.85\kpc$. The gas cells are adaptively softened according to the cell radius with a minimum value set to $0.85\kpc$. To explore the impact of reionization on low-mass galaxies, we have included simulations with \thesan physics replaced by a uniform UV Background (uniform UVB) model. In this model, we assume reionization is nearly instantaneous at redshift $z_{\rm i}=10$ (following the original IllustrisTNG model~\footnote{We note that the original Illustris and IllustrisTNG simulations adopted a fixed reionization redshift of $z_{\rm i}=6$. However, similar models like EAGLE initiate their UVB at a higher redshift of $z_{\rm i}=11.5$ \citep{Schaye2015}. Here we choose $z_{\rm i}=10$ since our fully coupled RHD simulations end at $z=5$ to ensure efficient use of computational resources.}) with the strength of the radiation field given by \citet{FG2009}. Simulations with this UVB are performed in exactly the same way as the \thesan models. The ionizing radiation flux from the UVB is then passed directly to the thermochemistry module. In most cases, it fully ionizes the gas, though gas can be self-shielded from such radiation if it reaches a high enough density. We employ the model from \citet{Rahmati2014} to determine if such gas is self-shielded against the external UVB. 

The fiducial \thesanhr runs can be viewed as putting the small volume in a void of galaxies since no external ionizing photon background is included. On the contrary, the uniform UVB model is equivalent to putting the small volume in a bath of ionizing photons contributed by luminous galaxies nearby. So we effectively bracket the uncertainties from reionization history/morphology in this experiment. This fact is also illustrated in the top panel of Figure~\ref{fig:sfrd-box}. The two types of runs have systematically faster/lower reionization history compared to the main \thesan suite which has been calibrated to reionization on large scales. A subtle point is that the instantaneous switch-on of the uniform UVB can generate an artificial bump in the star formation history (SFH) of galaxies at $z\gtrsim 10$. This population of old stars should not exist in the physical case when the volume is affected by ionizing feedback from massive galaxies.

\begin{figure}
    \centering
    \includegraphics[width=1\linewidth]{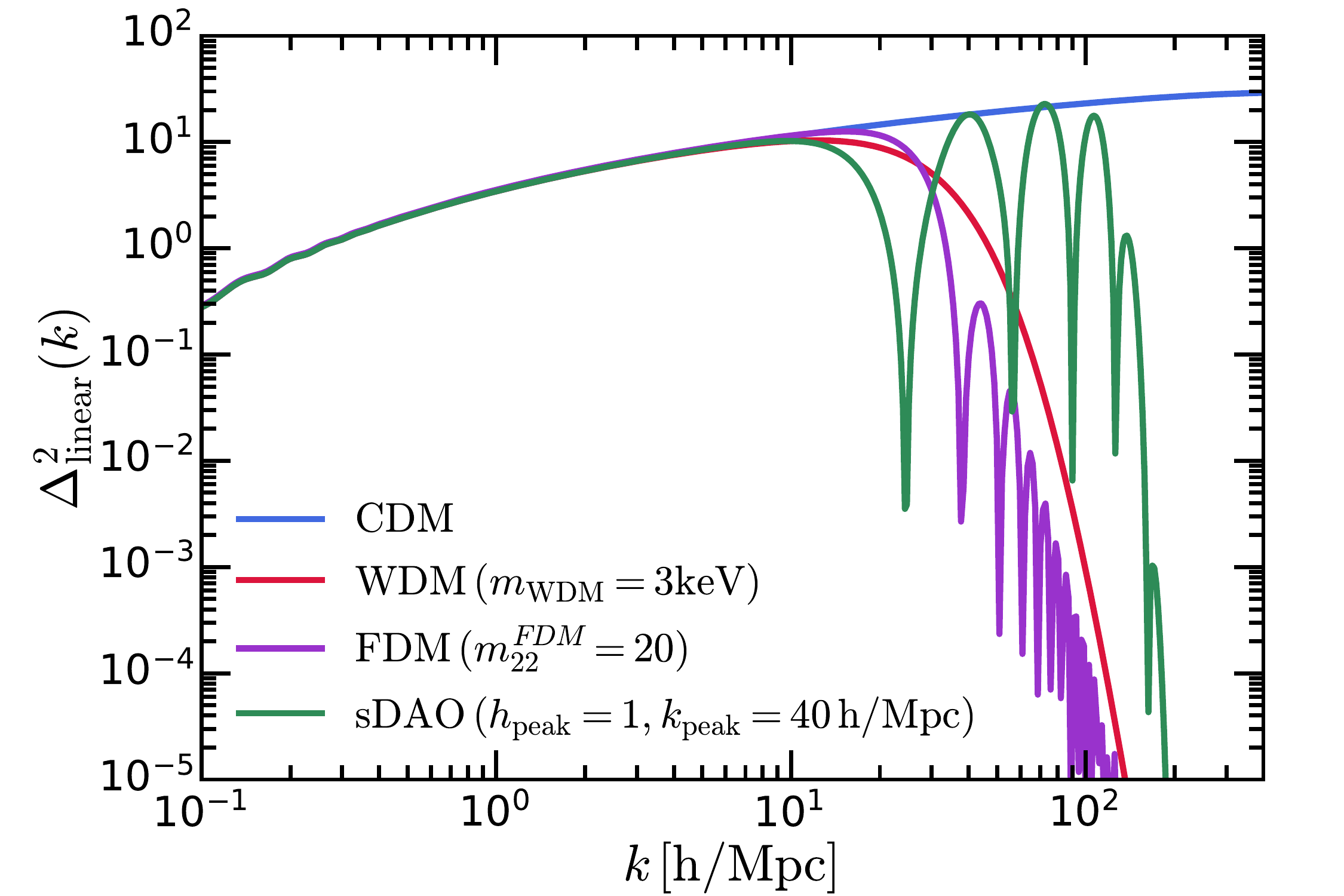}
    \caption{Linear matter power spectrum for the CDM (blue curve), WDM ($m_{\rm WDM}=3\kev$; red curve), sDAO (green curve), and FDM ($m_{\rm a}=2\times 10^{-21}\,{\rm eV}$; purple curve) models. $\Delta^2(k) \equiv k^{3}\,P(k)/2\pi^2$ is the dimensionless power spectrum. For the CDM and FDM models, the linear matter power spectrums are generated using {\sc Camb} and {\sc AxionCamb}. For the WDM and sDAO models, we employ analytical transfer functions (see Section~\ref{sec:sim-ic-dm} for details) on top of the CDM power spectrum. The altDM models examined in this study all have small-scale damping of the power spectrum. They differ in the location and shape of the damping wing as well as the oscillatory patterns.}
    \label{fig:power-spectrum}
\end{figure}

\subsection{Initial conditions and DM models}
\label{sec:sim-ic-dm}

The initial conditions are generated with the {\sc Gadget-4} code \citep{Springel2021} using the second-order Lagrangian method at the initial redshift $z=49$. For the ${\rm CDM}$ model, we adopt the linear matter power spectrum generated using {\sc Camb} \citep{CAMB1,CAMB2}, which is identical to the ones used in the fiducial \thesan simulations \citep{Kannan2022,Garaldi2022,Smith2022} and the IllustrisTNG simulation \citep{Nelson2019,Pillepich2018}. Notably, in addition to the vanilla CDM model, the suite includes simulations in WDM, FDM, and sDAO models. These models introduce cutoffs to the linear matter power spectrum at small scales and could affect early structure formation in non-trivial ways. The difference between altDM models and CDM is completely characterized by the transfer function, defined as \footnote{In this paper, we are interested in the main impact of deviations over CDM in the primordial power spectrum at (sub)galactic scales. Therefore, we have ignored quantum pressure and wave interference effects of FDM \citep[e.g.][]{Mocz2017,Nori2019,Mocz2020}. We also ignored the potential self-interactions of DM often associated with the DAO models \citep[e.g.][]{Cyr-Racine2014}, which has little impact on structure formation in the first $\Gyr$.}
\begin{equation}
    T(k) \equiv \left( \dfrac{P_{\rm altDM}(k)}{P_{\rm CDM}(k)} \right)^{1/2}.
\end{equation}

For the WDM model, the free-streaming of DM particles damps the fluctuations of the density field at small scales. The transfer function for thermal relic WDM can be described as
\begin{equation}
    T(k) = \left( 1 + (\alpha k)^{\beta} \right)^{\gamma},
    \label{eq:transfer-wdm}
\end{equation}
where $\beta=2\nu$, $\gamma=-5/\nu$ and $\nu$ takes the value $1.12$ \citep[e.g.][]{Viel2005}. $\alpha$ controls the characteristic scale of the damping~\footnote{We have $\alpha = (1/k_{1/2})\left[ \left(1/\sqrt{2}\right)^{1/\gamma} - 1 \right]^{1/\beta}$ dervied by setting $T^{2}(k_{1/2}) = 1/2$.}. The half-power wavenumber (where the linear WDM suppression reaches $1/2$ in terms of matter power w.r.t. the $\Lambda$CDM case) can be approximated as~\citep{Viel2005,Viel2013}
\begin{equation}
    k^{\rm WDM}_{1/2} \simeq 22\, h \Mpc^{-1}\, \left( \dfrac{m_{\rm WDM}}{3\kev} \right)^{1.11}\, \left( \dfrac{\Omega_{\rm DM}}{0.25} \right)^{-0.11}\,\left( \dfrac{h}{0.7} \right)^{1.22},
\end{equation}
where $m_{\rm WDM}$ is the mass of the thermal relic WDM particle and $\Omega_{\rm DM} \equiv \Omega_{\rm 0} - \Omega_{\rm b}$ is the cosmological abundance of DM. We choose $m_{\rm WDM}= 3\kev$, which is of interest in explaining the 3.55 keV X-ray line from Galactic center~\citep[e.g.][]{Bulbul2014,Boyarsky2014,Adhikari2017} and lies at the edge of existing observational constraints~\citep[e.g.][]{Irsic2017,Gilman2020,Nadler2021}. The half-power wavenumber defined here is different from the half-mode wavenumber quoted in some literature. The half-mode wavenumber is defined as where the transfer function becomes $1/2$ w.r.t. the $\Lambda$CDM case and is larger than the half-power wavenumber.

For the IDM model featuring DAOs, we choose the model in the ETHOS framework \citep{Cyr-Racine2014} studied in \citet{Bohr2021}. The small-scale damping of the linear matter power spectrum is caused by interactions between DM and relativistic particles in the early Universe and is accompanied by DAOs. We use the parametrization outlined in Equation~(3) of \citet{Bohr2020}, which is an extension to Equation~(\ref{eq:transfer-wdm}). The transfer function in the DAO model is mainly controlled by two parameters, $k_{\rm peak}$ (the wavenumber of the first acoustic peak) and $h_{\rm peak}$ (the amplitude of the first acoustic peak). We take the parameter choice, $k_{\rm peak} = 40\,h \Mpc^{-1}$, $h_{\rm peak}=1$ in \citet{Bohr2020}. In this case, the half-power wavenumber of the first damping wing in the power spectrum is 
\begin{equation}
    k^{\rm sDAO}_{1/2} = k_{\rm peak}/2.5 \simeq 16\, h \Mpc^{-1}.
\end{equation}

For the FDM model, the quantum mechanical nature of ultralight ALPs moving with the Hubble flow will prevent clustering below the de Broglie wavelength and result in the damping feature at small scales. The transfer function can be approximated as \citep{Hu2000}
\begin{equation}
    T(k) \simeq \dfrac{{\rm cos}(x^{3})}{1+x^{8}},\,\, x \equiv 1.61\,\left( \dfrac{m_{\rm a}}{10^{-22}\,{\rm eV}}\right)^{1/18}\,\left(\dfrac{k}{k^{\rm eq}_{\rm J}}\right),
\end{equation}
where $m_{\rm a}$ is the ALP mass, $k^{\rm eq}_{\rm J} \simeq 9 \, (m_{\rm a}/10^{-22}\,{\rm eV})^{1/2} \Mpc^{-1}$ is the effective ``Jeans'' length at matter-radiation equality. The half-power wavenumber is \citep{Hu2000} 
\begin{equation}
    k^{\rm FDM}_{1/2} = 4.5\, (m_{\rm a}/10^{-22}\,{\rm eV})^{4/9} \Mpc^{-1}.
\end{equation}
To be more accurate, we compute the linear matter power spectrum of FDM with the Boltzmann code {\sc Axioncamb} \citep{Hlozek2015} (developed based on {\sc Camb}; \citealt{CAMB1,CAMB2}). We choose the ALP particle mass $m_{\rm a}=2\times 10^{-21}\,{\rm eV}$, which yields $k^{\rm FDM}_{1/2} \simeq 25\, h\Mpc^{-1}$ similar to the WDM model we study. The isocurvature fluctuations of FDM are ignored, which usually appear at sub-parsec scales~\citep[e.g.][]{Hogan:1988mp,Kolb:1993zz,Zurek:2006sy}. Following previous work on high-redshift structure formation in FDM \citep[e.g.][]{Schive2016, Armenguad2017}, we do not include the quantum pressure in the dynamic of our simulations. This should not alter our predictions for galaxies significantly as found in e.g. \citet{Zhang2018, Nori2019, Mocz2020}.

In Figure~\ref{fig:power-spectrum}, we show the linear matter power spectrum of all the models considered in this paper. The details and numerical parameters of the simulations are summarized in Table~\ref{tab:sims}. In the following parts of the paper, we will discuss the results of these simulations.

\begin{figure*}
    \raggedright
    \includegraphics[width=0.49\linewidth]{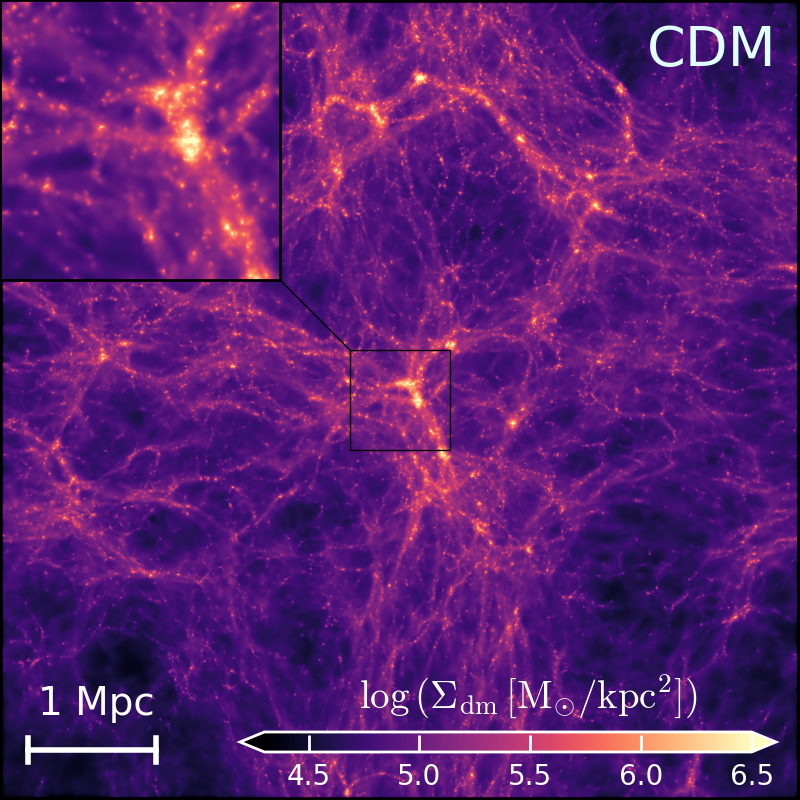}
    \includegraphics[width=0.49\linewidth]{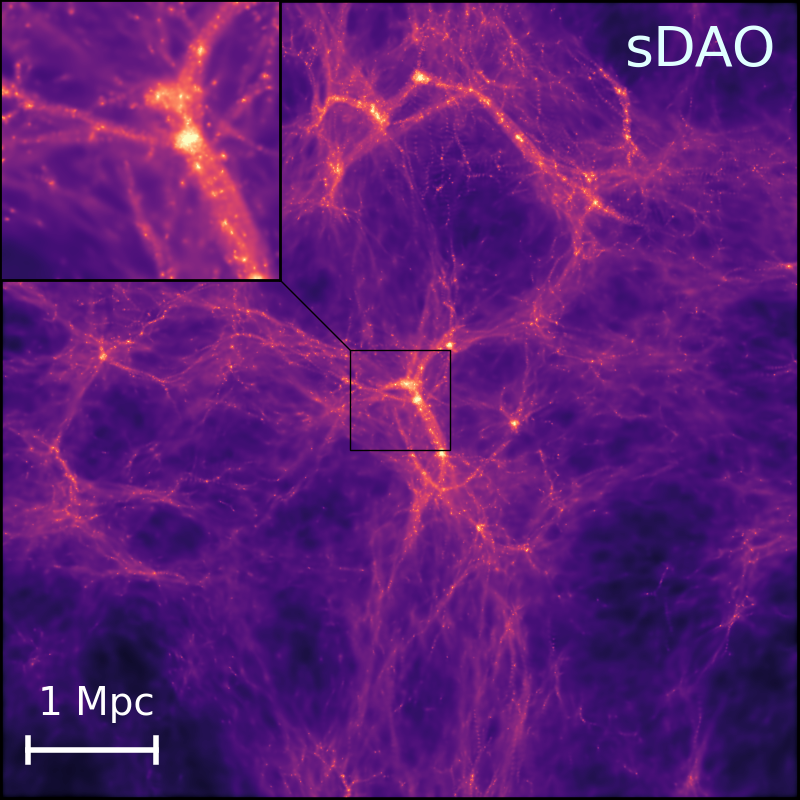}
    \includegraphics[width=0.49\linewidth]{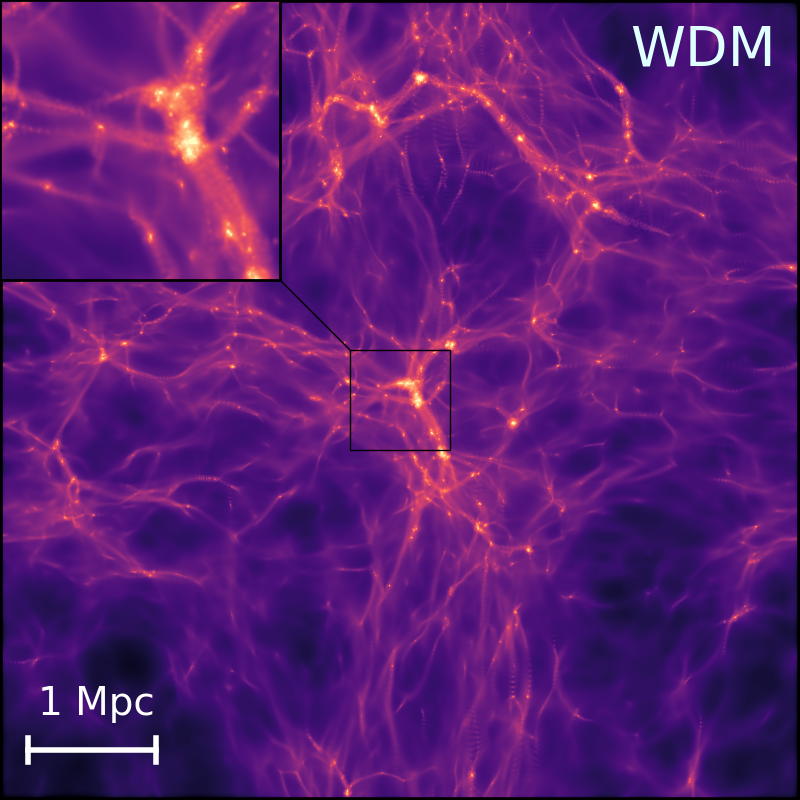}
    \includegraphics[width=0.49\linewidth]{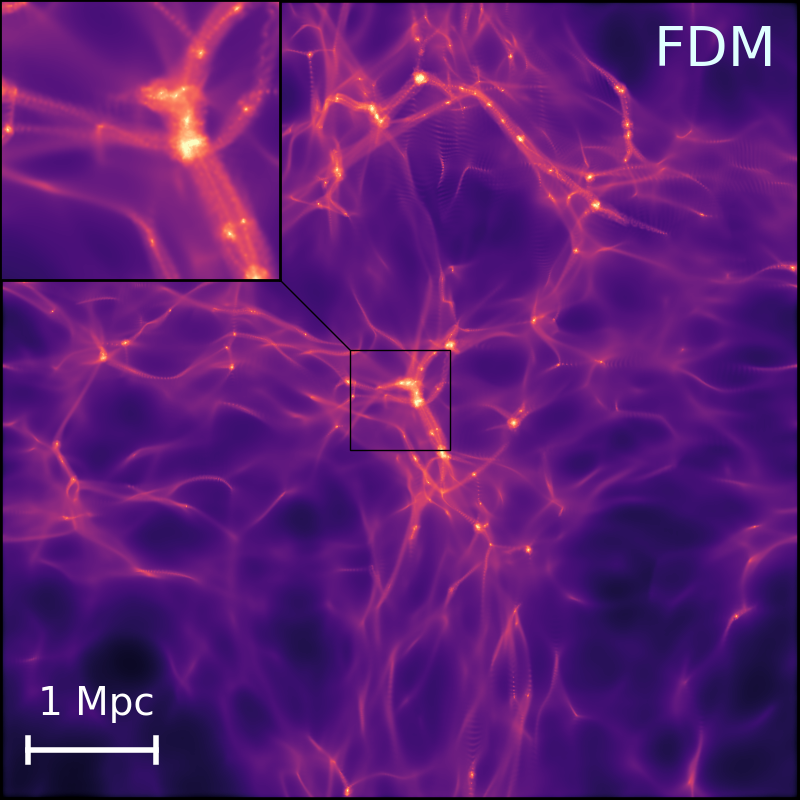}
    \caption{DM surface density map at $z=6$ centered on the most massive halo in the simulation volume. Here we show the runs with the fiducial \thesan physics (with RT). The DM density fields in runs with uniform UVB are identical to their RT counterparts. In the top left corner of each panel, we show a zoom-in image of the structures around the massive halo. Due to the suppressed small-scale power spectrum in altDM models, fewer low-mass haloes and filamentary structures formed. The sDAO model has residual fluctuations at small scales due to the DAOs. It has subtle differences from the WDM and FDM models with a single damping of the power spectrum. The overall density field in sDAO is smoothed out while a limited number of small-scale structures can still form.}
    \label{fig:image-dm}
\end{figure*}

\begin{figure*}
    \raggedright
    \includegraphics[width=0.49\linewidth]{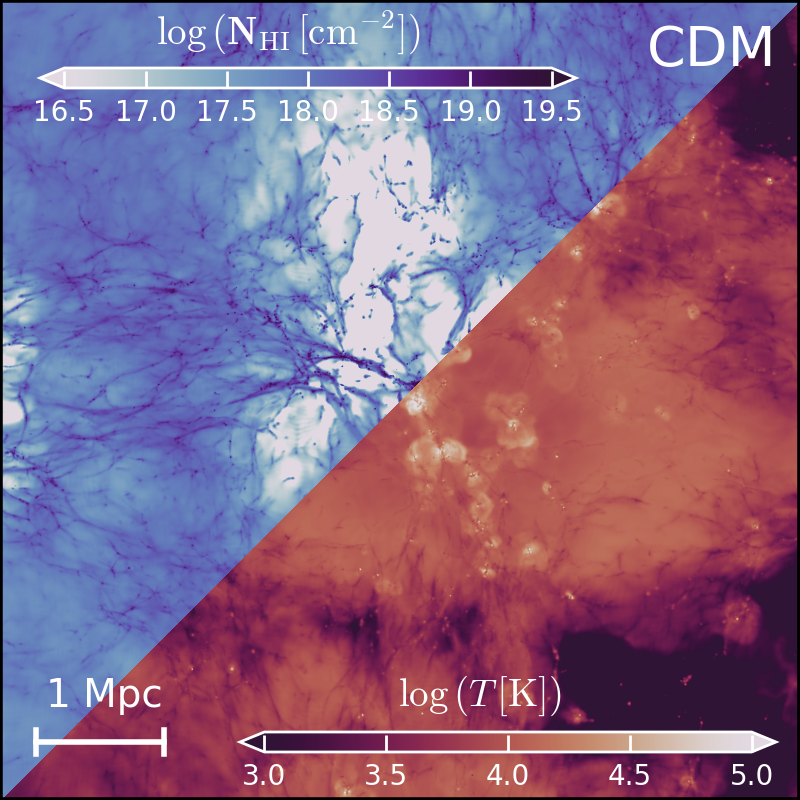}
    \includegraphics[width=0.49\linewidth]{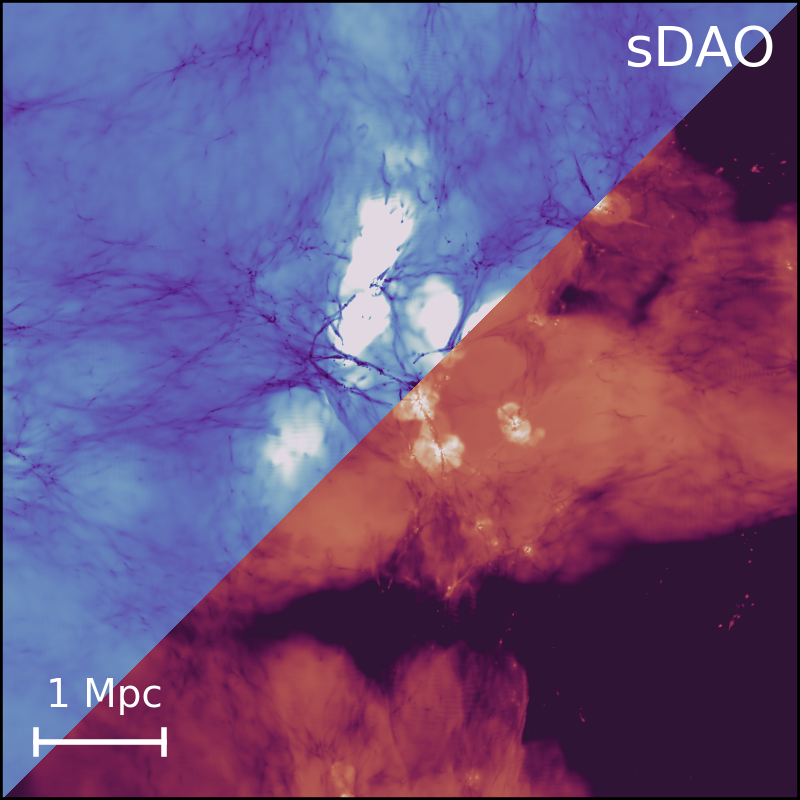}
    \includegraphics[width=0.49\linewidth]{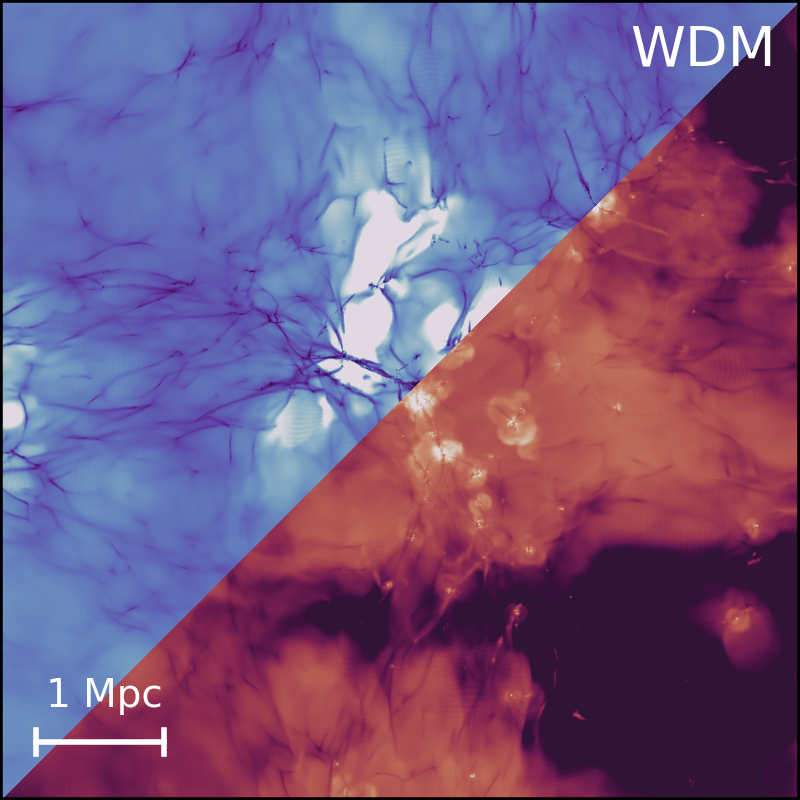}
    \includegraphics[width=0.49\linewidth]{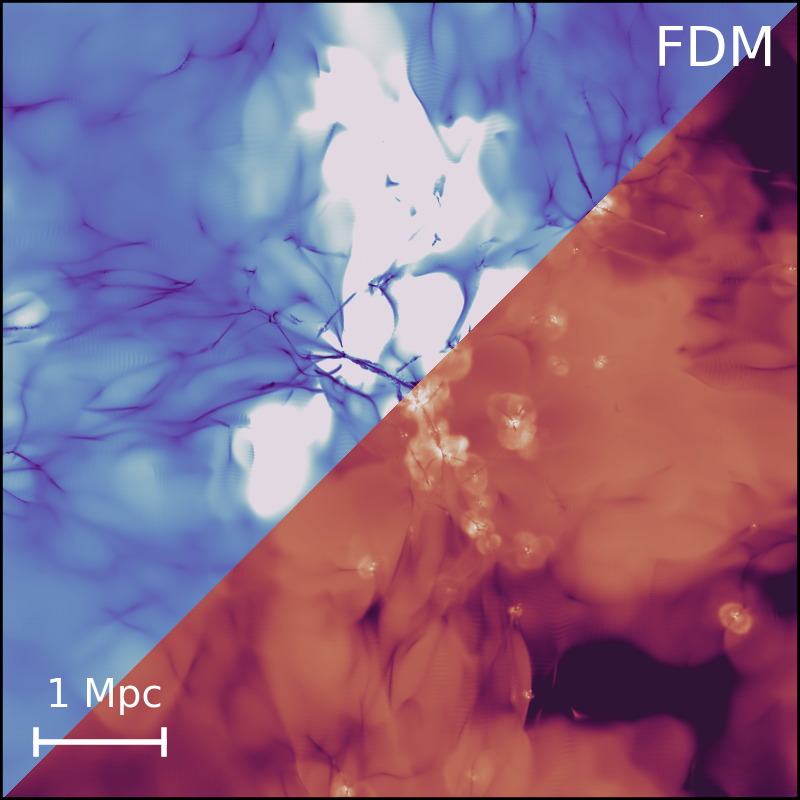}
    \caption{ Overview of the properties of the IGM in simulations at $z=6$. Each panel corresponds to one DM model studied. In each panel, the top left part shows the column density distribution of neutral hydrogen. The lower right part shows the (mass-weighted) gas temperature of the same volume as the top left part mirrored. The CDM model has more small clumps of neutral gas retained in subhaloes even in the highly-ionized regions. These low-mass subhaloes and associated cold gas clumps are missing in altDM models. At large scales, WDM and sDAO runs have a higher neutral gas abundance on average and smaller ionizing bubbles at $z=6$. The global reionization processes in these two models have been delayed systematically due to the deficiency of ionizing photons contributed by low-mass galaxies. Corresponding features can be found in the temperature distribution as well. The WDM and sDAO runs show larger voids of low-temperature, neutral gas while a lower number of small hot gas bubbles. In the FDM run, although the statistics of high-density gas clumps, filaments, and hot gas bubbles are similar, we find rather different phenomena on large scale. The global ionization fraction at $z=6$ has overtaken the CDM value due to the late-time starburst (shown in Figure~\ref{fig:sfrd-box} as well). These strong non-linear effects create larger ionizing bubbles and lower neutral gas column densities in FDM at $z\lesssim 6$.}
    \label{fig:image-gas}
\end{figure*}

\section{DM density field and IGM properties}
\label{sec:visual}

The suppression of the linear matter spectrum is directly reflected in the DM density field. In Figure~\ref{fig:image-dm}, we show images of the DM surface density distributions at $z=6$ in different models listed in Table~\ref{tab:sims}. The density map is projected through a slice of thickness $\sim 3\,{\rm cMpc}$ (roughly half of the box size). Due to the suppression of the power spectrum, the altDM models give smoother DM distributions on small scales and a lower number of filaments and low-mass (sub)haloes. The smoothing of the density field is most noticeable for the FDM model with the steepest damping of the matter power spectrum. On the contrary, the sDAO model with acoustic peaks at large wavenumber preserves some small-scale features. The DM density fields shown in this Figure~\ref{fig:image-dm} are obtained from runs utilizing the fiducial \thesan physics. Runs with uniform UVB yield almost the same DM distribution and are shown in the Appendix~\ref{app:visual}.

\begin{figure}
     \centering
    \includegraphics[width=1\linewidth]{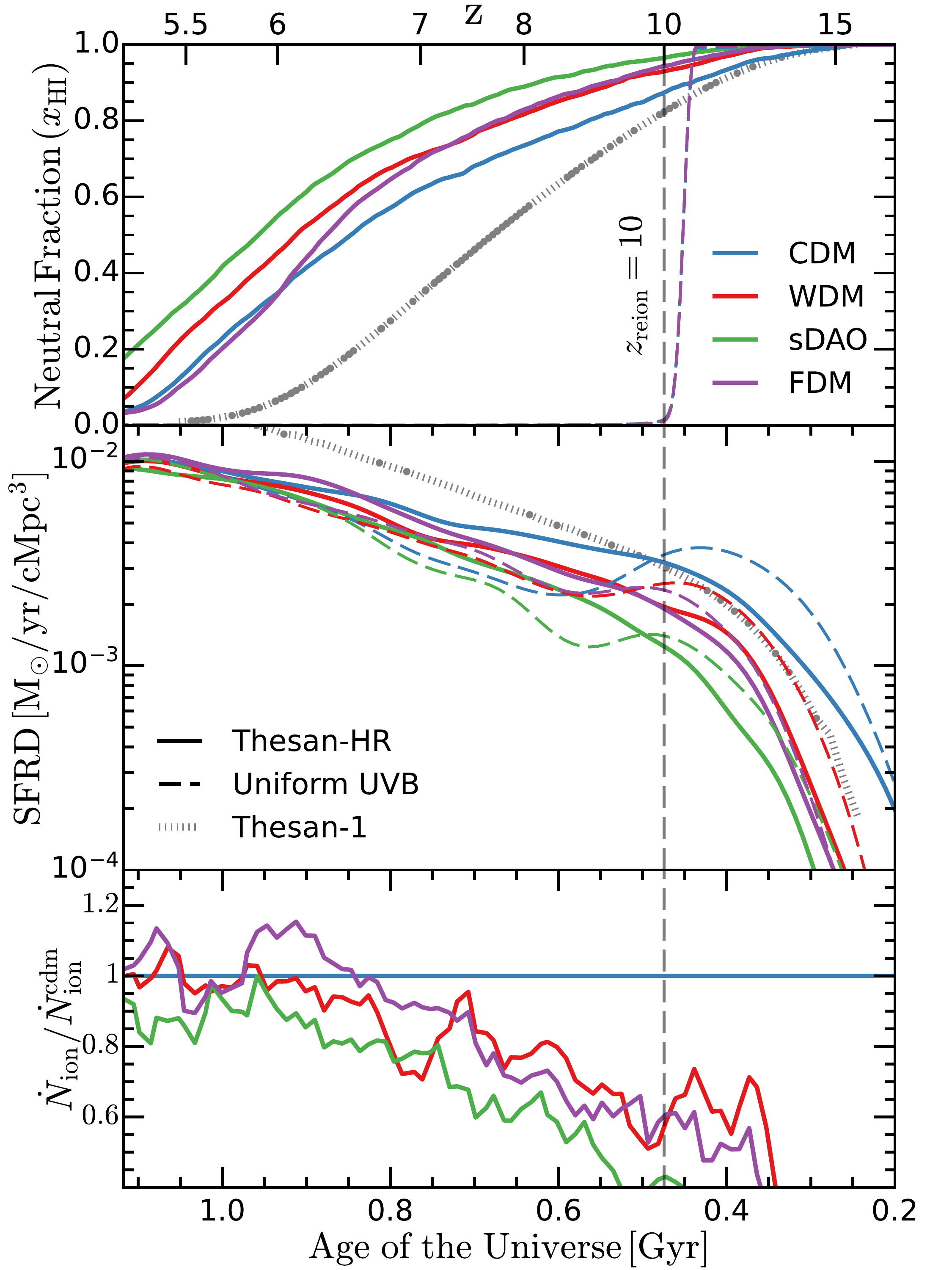}
    \caption{{\itshape Top:} Evolution of the global volume-weighted neutral hydrogen fraction in simulations. The solid lines show the results from the \thesanhr simulations with RT, while the dashed lines show the results in the uniform UVB model. In altDM models, reionization histories display systematic delays that correlate with the damping scale of the power spectrum. Interestingly, in the FDM model with the sharpest power spectrum damping, late-time starburst in low-mass galaxies can lead to a global ionized fraction larger than the CDM counterpart at $z\lesssim 6$. {\itshape Middle:} Evolution of the star formation rate density (SFRD) in the simulation volume. Star formation in the altDM models is suppressed at high redshift due to delayed structure formation at small scales until the late-time starburst closes the gap. In contrast, the uniform UVB model exhibits a bump in SFRD before the activation of the uniform UVB. After the UVB is turned on of the UVB, star formation significantly decreases. {\itshape Bottom:} Evolution of ionizing photon production rate with respect to the value in CDM. Only the results of RT simulations are shown. The late-time starburst in altDM models quickly closes the gap in ionizing photon production. The $\dot{N}_{\rm ion}$ in FDM surpasses the value in CDM at $z \sim 7$ which drives the acceleration of reionization afterward.
    }
    \label{fig:sfrd-box}
\end{figure}

The suppression of small-scale DM structures will delay the formation of low-mass galaxies, which are the dominant sources of ionizing photons in the early phase of reionization \citep[e.g.][]{Jaacks2012,Madau2014,Robertson2015,Rosdahl2022,Yeh2023}. Therefore, the morphology of reionization and the thermal properties of the IGM can be affected. In Figure~\ref{fig:image-gas}, we show the column density map of neutral hydrogen at $z=6$ in the simulations. The camera position and the slice of volume visualized are identical to the DM image (Figure~\ref{fig:image-dm}). The neutral gas in the IGM is a highly-biased tracer of the underlying DM distribution \citep[e.g.][]{Furlanetto2006,Pritchard2012,Rahmati2013,Navarro2018}. The high-density neutral gas (often in the form of dense self-shielded clumps) represents baryons falling into gravitationally bound DM structures and thus traces DM overdensities well. Some less dense neutral gas is retained in filaments around DM haloes, which will be fed into DM haloes fueling subsequent star formation. These two types of neutral gas structures are more abundant with increasing clustering power of DM at small scales and are suppressed in altDM models. On the other hand, the low-density neutral gas filling the IGM is more sensitive to the phase of global reionization that ties to the global SFH. In the WDM and sDAO models, the star formation and reionization history are systematically delayed compared to the CDM case (as will be shown in Figure~\ref{fig:sfrd-box}). Therefore, the overall opacity of neutral hydrogen on large scales is enhanced, contrary to the trend at high densities and small scales. In the FDM model, it is the same situation at $z\gtrsim 7$. But between $z=6$ and $z=7$, the neutral gas preserved by the delayed early-phase reionization leads to enhanced star formation in low-mass galaxies. It elevates the total star formation in the simulation volume (despite lowered number counts of faint galaxies) and accelerates the late-time reionization to even overtake the CDM counterpart. This will be shown in Figure~\ref{fig:sfrd-box} and discussed below. The decreased abundance of dense neutral clumps and filaments at small scales allows the hot ionizing bubbles to penetrate through the IGM in a more violent fashion. This is why the IGM at $z=6$ in the FDM model is less opaque than the CDM case. Overall, the IGM properties are more sensitive to the location of the first damping in the matter power spectrum than the small-scale acoustic oscillations. The sDAO model with strong small-scale oscillations ($h_{\rm peak}=1$) and steeply rising halo mass function at the low-mass end (compared to WDM and FDM) still has the most delayed star formation and reionization history out of the three altDM models.

The gas temperature map is also shown in Figure~\ref{fig:image-gas}. Bubbles of gas are heated by stellar feedback and are spatially correlated with DM haloes that are massive enough to host galaxy formation. In the WDM and sDAO runs, we find fewer small-scale clusters of these hot ionized bubbles and larger voids of cold neutral gas on large scales compared to CDM. In the FDM model, due to the strong late-time star formation we mentioned above, the volume-filling fraction of hot ionized bubbles is larger than in the CDM case. The radiative heating of the IGM is dominated by large bubbles from massive galaxies. In general, the temperature distribution of IGM gas is consistent with the trends observed in the neutral gas distribution. The IGM properties in the uniform UVB model are shown in the Appendix~\ref{app:visual}. In short, averaging the ionizing photon background of luminous sources over the entire simulation volume ionizes all of the gas. This produces radically different morphologies of reionization and thermal properties of the IGM. The entire volume is heated to above $\sim 10^{4}\,{\rm K}$.

In the top panel of Figure~\ref{fig:sfrd-box}, we show the volume-weighted neutral fraction of hydrogen in simulations. Regardless of the DM model, runs with the uniform UVB have the entire simulation volume ionized right after the activation of the UVB. The \thesanone result is shown for reference with the dotted line. It emphasizes that the small-volume \thesanhr runs are tracing a biased volume of the Universe. This bias is mainly caused by the absence of massive, bright sources of ionizing photons that would be present in a real environment. {\itshape Therefore, we will be cautious about interpreting our results quantitatively as what would be found in real observations but focus on the net differences driven by various of physics effects.} 

All altDM models cause a delay in cosmic reionization during the early phase when low-mass galaxies primarily contribute to ionizing photons. The extent of the delay in reionization is positively correlated with the damping length/mass scale of the power spectrum. At late stages of reionization ($z \lesssim 7$), the FDM model exhibits a period of enhanced star formation and accelerated reionization which enables it to surpass the CDM at $z\lesssim 6$. However, it is important to note that the phenomena observed in small-volume simulations may not necessarily be representative of the global reionization signal. For example, at $z\sim 6$, the SFRD and the ionizing photon budget on $\sim 100\Mpc$ scales are predominantly influenced by massive galaxies ($M_{\ast}\gtrsim 10^{9}\msun$, e.g. \citealt{Kannan2022,Yeh2023}), which are not sampled in \thesanhr. Therefore, the signal in the global phase of reionization will be much weaker.

In the middle panel of Figure~\ref{fig:sfrd-box}, we show the redshift evolution of the star formation rate density (SFRD) in the entire simulation volume. This is derived from the initial-mass-weighted age distribution of all stellar particles collected at $z=5$. Again, it is important to note that the results should not be compared to a large-volume-averaged ``global'' cosmic SFRD due to the limited simulation box size. In the WDM and FDM models, star formation is delayed by approximately $100\Myr$, while in the sDAO model, the delay is around $200\Myr$. Star formation in these altDM models can quickly ``catch up'' to the CDM case through a strong late-time starburst. This phenomenon will be discussed in more detail in Section~\ref{sec:uvlf} and Section~\ref{sec:sfrd}. The uniform UVB model exhibits a bump in star formation at $z \sim 12$ before the uniform UVB is activated. Star formation at $z\lesssim 10$ is suppressed as the neutral gas in the diffuse IGM is rapidly ionized by the background radiation. Artificial signals, such as WDM/FDM overtaking CDM at $z\sim 8$, can show up since they consume less gas in the early bump of star formation.

In the bottom panel of Figure~\ref{fig:sfrd-box}, we present the redshift evolution of the intrinsic ionizing photon production rates, denoted as $\dot{N}_{\rm ion}$, normalized by the values in CDM. Within altDM models, the production of ionizing photons is suppressed at high redshifts compared to CDM. At lower redshifts, this disparity narrows due to the strong late-time star formation in altDM models. This trend resembles our findings for the SFRD. Notably, the FDM model surpasses CDM in ionizing photon production rate at $z\lesssim 7$, which marks the onset of a period of accelerated reionization as shown in the top panel.

\section{Halo and stellar mass functions}
\label{sec:hmf-smf}

\begin{figure}
    \centering
    \includegraphics[width=1\linewidth]{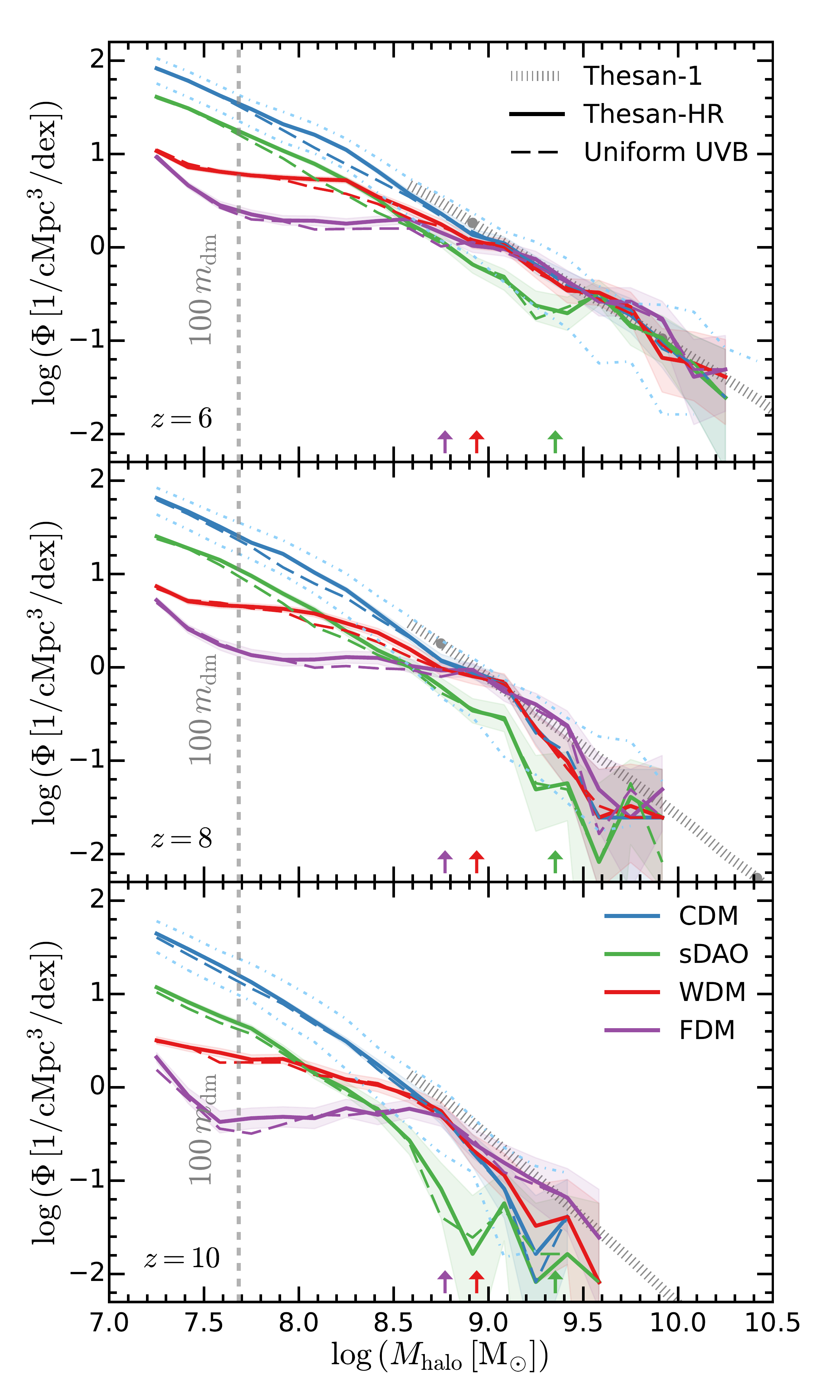}
    \caption{Halo mass function at $z=$6, 8 and 10 from top to bottom. We show the halo mass functions from the simulations in different DM models. For each model, results from the \thesan model (the uniform UVB model) are shown as solid (dashed) lines. The faint blue dotted-dashed lines show the cosmic variance estimated in CDM. The halo mass function in the \thesanone simulation is shown as the black dotted line. The arrows at the bottom of each panel indicate the half-power mass of altDM models. Below the half-power mass, we find suppression to the halo mass function due to altDM physics, of which the shape depends on the shape of the damping feature in the power spectrum.} 
    \label{fig:hmf}
\end{figure}

The low-mass end of the (sub)halo mass function is directly affected by the suppression of the small-scale power spectrum in altDM models. In Figure~\ref{fig:hmf}, we show the halo mass function in different models at $z=6,8,10$. The $1\sigma$ confidence intervals of number densities are estimated using the \citet{Gehrels1986} formulae and are shown as the shaded regions. Throughout this paper, the ``halo'' is defined as gravitational bound structures identified by the {\sc Subfind} algorithm \citep{Springel2001}, including both central and satellite galaxies. Halo mass is defined as the total mass of all particles (or cells) gravitationally bound to the halo. The high resolution achieved by the \thesanhr suite allows us to make reliable predictions of halo properties down to about $5\times 10^{7} \msun$ (as indicated by the vertical dashed line in the figure). This is close to the atomic cooling limit, $\sim 10^{7.5-8}\msun$ \citep{Wise2014}, although we do not explicitly model the low-temperature molecular phase of gas and the Lyman-Werner (LW) radiation that dissociates molecular hydrogen. 

All altDM models show suppression of the halo mass function in the low-mass end. In Appendix~\ref{app:half-power}, we estimate the characteristic mass scale of this suppression according to the half-power wavenumber $k_{1/2}$ in the linear matter power spectrum. The half-power halo mass $\log{(M_{\rm halo}(k_{1/2})/\msun)}$ is $8.9$, $8.8$, and $9.4$ for the WDM, FDM, and sDAO models studied in this paper, respectively. They accurately predict where the suppression in halo mass function shows up. In addition, different from the cut-off-like feature in WDM and FDM, the sDAO model predicts a steep rise of halo mass function in the low-mass end similar to CDM but with lower normalization. As discussed in \citet{Bohr2021}, this is due to the strong DAOs in this model (with the first acoustic peak retaining $100\%$ of the CDM power), which preserve some small-scale fluctuations of the primordial density field. It is known for the WDM and FDM models (or in general any model with steep suppression of power spectrum), spurious structures can form at the limiting mass $M_{\rm lim} = 10.1 \bar{\rho}_{\rm m}\,(L_{\rm box}/N_{\rm dm}^{1/3})\,k^{-2}_{\rm p}$~\citep{Wang2007}, where $k_{\rm p}$ is where the dimensionless power spectrum ($\Delta^{2}(k)$) reaches its maximum which is about $10\,h/\Mpc$ for the WDM and FDM models studied here. This limiting mass is calculated to be $\log{(M_{\rm halo}/\msun)} \simeq 7.8$, which is consistent with the scale of the uptick in the FDM run in Figure~\ref{fig:hmf}. This limiting mass is comparable to the mass of the smallest resolved haloes in the simulation (defined to be haloes of mass equivalent to 100 DM particles). Therefore, in the rest of the paper, we will not consider the problems caused by spurious haloes.

Cosmic variance is important for these small-volume simulations. We estimate this effect by subsampling $L=4\Mpc/h$ volumes in the \thesanhrlarge run (listed in Table~\ref{tab:sims}) with the same resolution as \thesanhr runs but larger box size, $L_{\rm box}=8\Mpc/h$. The variance in halo mass function obtained through this subsampling is then multiplied with the variance inflation factor computed in Appendix~\ref{app:cosmic-var}. The cosmic variance in CDM is shown with the blue dotted dashed lines in Figure~\ref{fig:hmf}. The observed differences between DM models are statistically robust against the cosmic variance.

\begin{figure}
    \centering
    \includegraphics[width=1\linewidth]{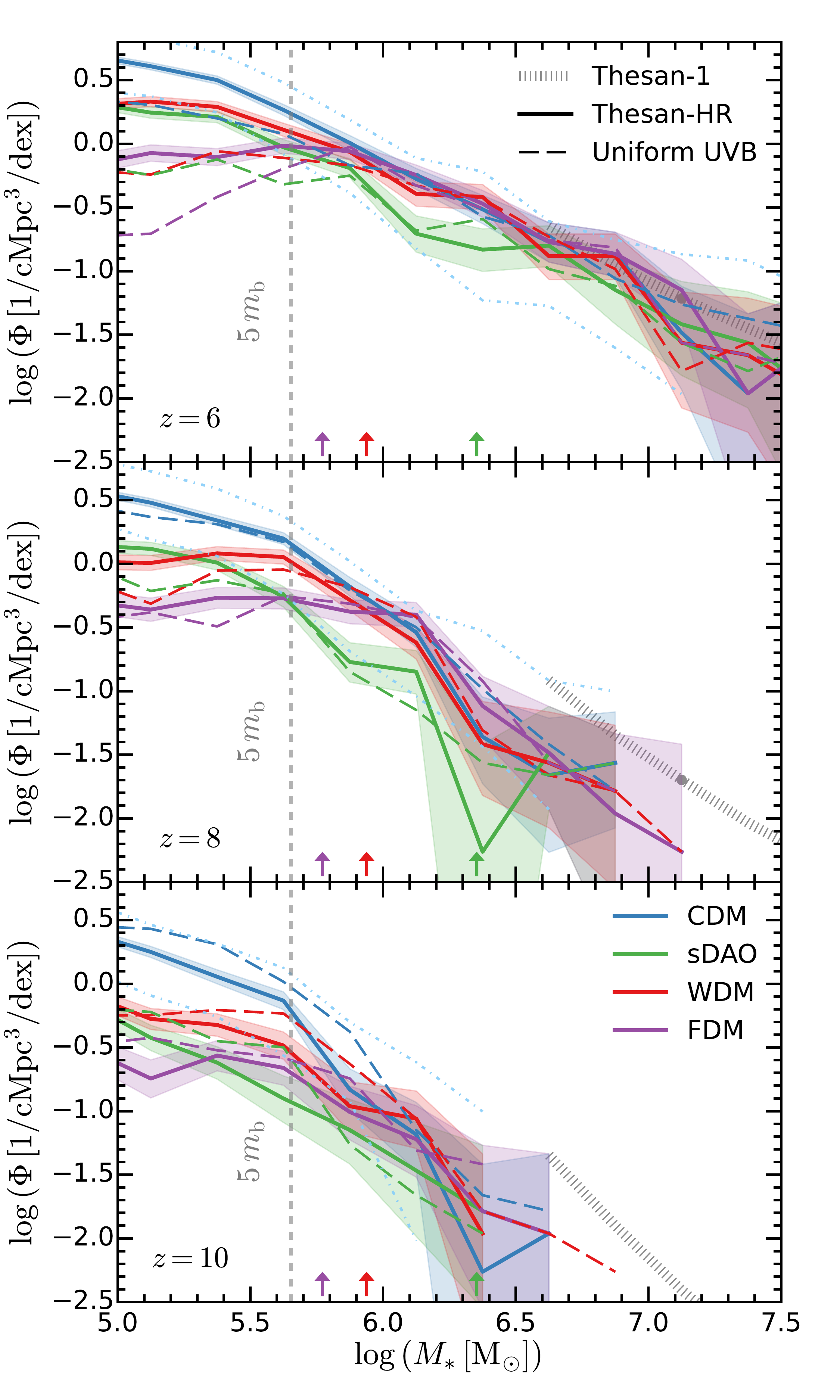}
    \caption{Galaxy stellar mass function at $z=$6, 8 and 10 in different models. The labeling follows Figure~\ref{fig:hmf}. The vertical lines indicate the mass roughly corresponding to $5$ stellar particles in the galaxy. Similar to the halo mass function, we find suppression at the low-mass end of the stellar mass functions in altDM models below the half-power stellar mass. The signature is less prominent due to increased statistical noises and cosmic variance. The uniform UVB model can suppress the stellar mass function in a similar fashion to altDM after the activation of the UVB.}
    \label{fig:smf}
\end{figure}

In Figure~\ref{fig:smf}, we show the stellar mass functions in different models at $z=6,8,10$. The stellar mass of a galaxy is defined as the total mass of stellar particles within $R_{\rm max}$~\footnote{$R_{\rm max}$ is the radius where the circular velocity $\equiv \sqrt{GM(<r)/r}$ of the halo reaches its maximum. For an NFW profile, this is directly related to the virial radius as $R_{\rm max} = 2.16\,R_{\rm vir}/c$, where $c$ is the concentration parameter with typical values of $4$ (insensitive to halo mass at high redshift, e.g. \citealt{Bullock2001}). The definition we used here is different from what most literature used. The main reasons for picking $R_{\rm max}$ as the aperture are: (1) It is less biased. The internal DM distribution in haloes is less affected by altDM physics despite slightly lowered concentration \citep[e.g.][]{Lovell2014,Bose2016-CoCo,Bohr2021}. We avoid potential indirect influence on physical quantities coming from aperture definitions. (2) $R_{\rm max}$ is a metric directly available in the {\sc Subfind} catalog. We have compared it to classical definitions e.g. stellar mass within twice the stellar-half-mass radius of the galaxy, and they agree reasonably well with no signs of systematical differences between DM models.}. In low-mass galaxies, due to the steeply decreasing stellar-to-dark-matter-mass ratio, the statistics of stellar particles are much poorer than the DM counterparts. We mark five times the baryonic mass resolution of simulations as the vertical dashed line in the figure. The Poisson sampling noise is shown with the shaded region. We estimate the cosmic variance (in CDM) following the same subsampling method used for the halo mass function, and it is illustrated using blue dotted-dashed lines. They are larger compared to those in halo mass functions due to lowered number of halos that host star formation and additional variances from galaxy formation physics.

For altDM models, the suppression of the abundance of low-mass haloes is reflected in the stellar mass function, although to a less extent. For instance, a reduction of approximately $0.5\,{\rm dex}$ is observed at $M_{\ast} \lesssim 10^{6}\msun$ at $z=6-10$ for the sDAO model. The arrows indicate the estimated half-power stellar masses of altDM models, derived using the methodology detailed in Appendix~\ref{app:half-power}. The mass scale of the suppression is consistent with these estimations. However, the signatures of altDM in the stellar mass functions are less prominent compared to those in the halo mass function. They are also less pronounced at lower redshift. Part of the reason is limited statistics (Poisson noise) and additional variance introduced by galaxy formation physics. Another important factor is that galaxies in altDM models are forming stars more rapidly at late times to ``catch up'' with CDM, as will be discussed in more detail in the following sections. 

The uniform UVB model results in the suppression of the stellar mass function at $M_{\ast} \lesssim 10^{6}\msun$ at $z=6$. This coincides with where altDM signatures become apparent. Once the uniform UVB is activated, reionization occurs quickly throughout the entire simulation volume, with the exception of a few dense, self-shielded clumps in massive galaxies. This rapid reionization suppresses star formation in low-mass galaxies. However, at redshift $z=10$, the stellar mass function displays an increase at low masses. Prior to the activation of the UVB, the ignorance of local ionizing sources leads to enhanced star formation. This corresponds to the artificial bump of star formation at $z\gtrsim 10$ in the global SFH, as shown in Figure~\ref{fig:sfrd-box}.

\section{Rest-frame UV luminosity function}
\label{sec:uvlf}

\begin{figure}
    \centering
    \includegraphics[width=1\linewidth]{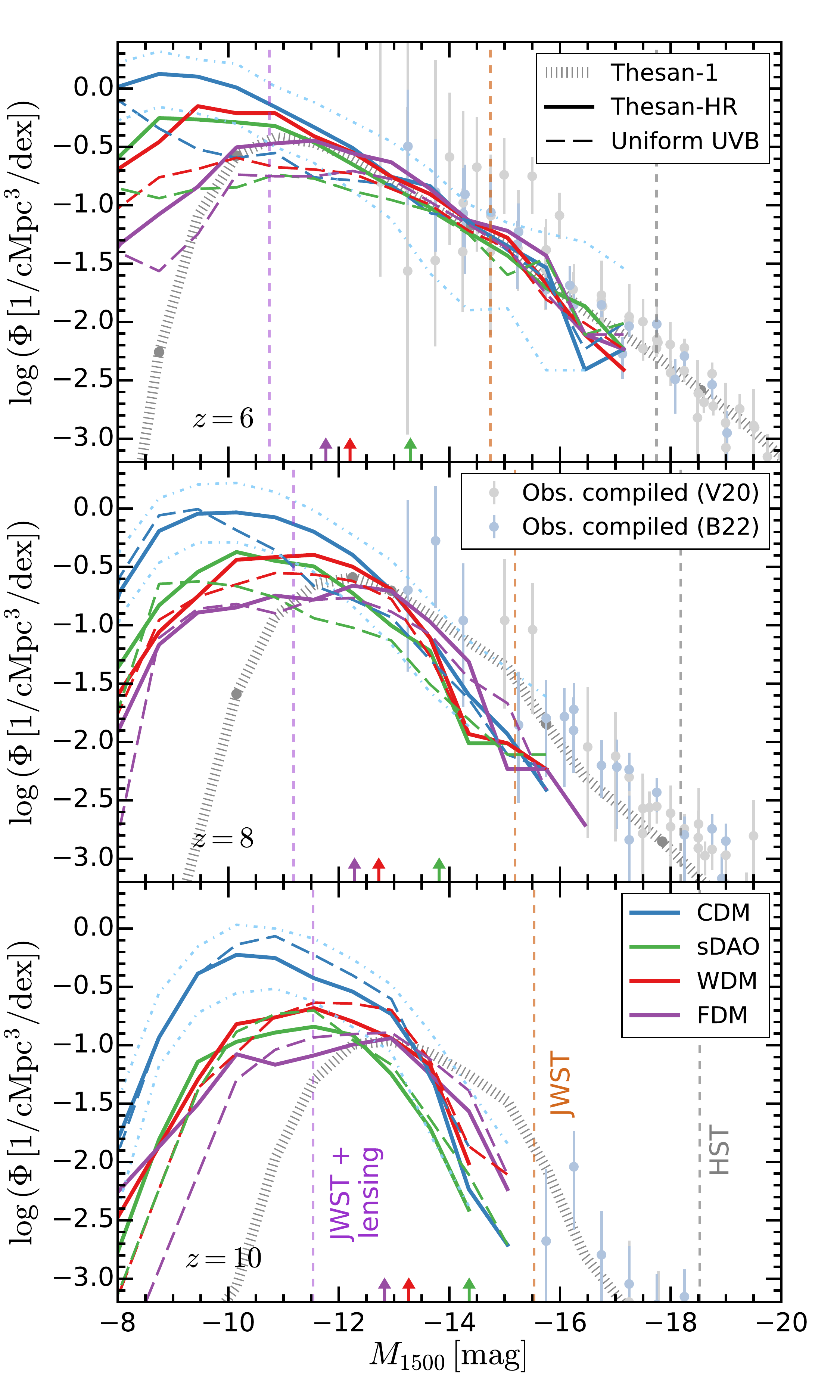}
    \caption{Rest-frame UV luminosity function at $z=6,8,10$. The UV luminosities are calculated using the population synthesis code and an empirical dust correction. The cosmic variance effect is illustrated with the dotted-dashed line. The observational constraints are taken from the compilation in \citet{Vogelsberger2020} and \citet{Bouwens2022} (see the main text for details). The vertical lines show the detection limits of HST and JWST. For reference, the UV luminosity function from the \thesanone simulation is shown as the black dotted line. On the relatively bright end ($M_{\rm 1500} \lesssim -13$), \thesanone prediction as well as the small-volume simulations at $z=6$ are in good agreement with the observations. Comparing different DM models, the suppression at the faint end due to delayed structure formation at small scales is apparent and similar to what we found in the stellar mass functions. They could be revealed by JWST observations of lense-magnified systems.}
    \label{fig:rest_uvlf}
\end{figure}

To better illustrate the detectability of altDM signatures, in Figure~\ref{fig:rest_uvlf}, we show the rest-frame UV luminosity functions of galaxies in the simulations. The rest-frame UV (at $1500$ \AA) luminosities of galaxies are calculated using the Binary Population and Spectral Synthesis models (BPASS v2.2.1; \citealt{Eldridge2017}), which is consistent with the choice of the \thesan model in computing the ionizing photon spectrum. A correction is then applied for galaxies with poor sampling of SFHs (see Section 3 of \citealt{Smith2022}). Dust attenuation is approximated using the formula in \citet{Gnedin2014} assuming the dust-to-metal ratios calibrated in \citet{Vogelsberger2020} (although dust attenuation has negligible impact on the faint-end of the UV luminosity function). 

The luminosity functions are compared to that of \thesanone \citep{Kannan2022} and the latest observational constraints. We include the observational data compiled in \citet{Vogelsberger2020}, including results from \citet{Mclure2013, Schenker2013, Oesch2013, Bouwens2015, Finkelstein2015, Livermore2017, Bouwens2017, Atek2018, Oesch2018}. In addition, we include new observational constraints from the HFF compiled in \citet{Bouwens2022}, including results from \citet{Ishigaki2018, Bhatawdekar2019, Bouwens2022}. The observational limit for the HST and JWST are derived following \citet{Jaacks2019}, assuming limiting AB magnitudes of $m^{\rm lim}_{\rm band} \sim 29$ and $32$, respectively. If neglecting IGM absorption, we can convert them to rest-frame UV magnitudes as
\begin{equation}
    M^{\rm rest}_{\rm UV} = m_{\rm band} + 2.5 \log{\left( \dfrac{1+z}{(D_{\rm L}(z)/10\pc)^{2}}\right)},
\end{equation}
which roughly gives $M^{\rm rest}_{\rm UV} = m_{\rm band} - 46.7$ at $z=6$, $M^{\rm rest}_{\rm UV} = m_{\rm band} - 47.2$ at $z=6$ and $M^{\rm rest}_{\rm UV} = m_{\rm band} - 47.5$ at $z=10$. For lensed fields with JWST, we assume it is $5\mmag$ deeper than the unlensed field (corresponding to a typical lens-magnification factor $\sim 100$, e.g. \citealt{Ishigaki2018, Bouwens2022}). 

\begin{figure*}
    \centering
    \includegraphics[width=1\linewidth]{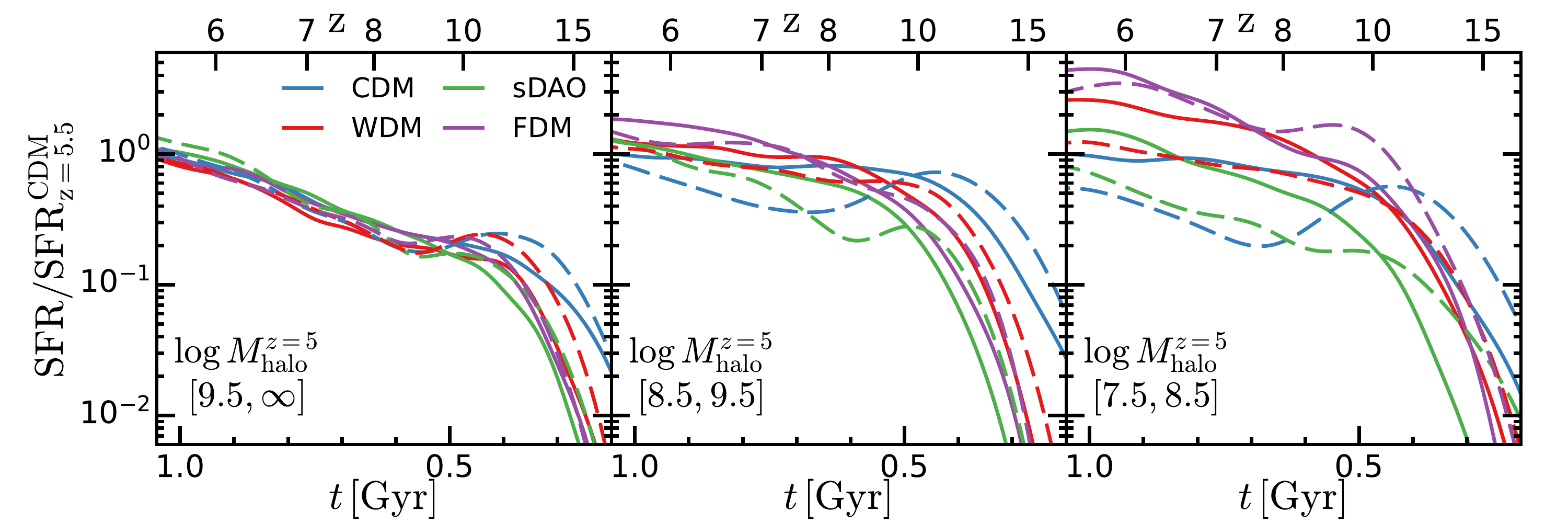}
    \caption{Average SFH of haloes selected at $z=5$ binned by halo mass. We choose three mass bins $\log{(M^{z=5}_{\rm halo}/\msun)} \in [7.5,\,8.5]$, $[8.5,\,9.5]$ and $[9.5, \infty]$. In the low-mass and intermediate-mass bins, we find patterns that are consistent with the evolution of the overall SFRD shown in Figure~\ref{fig:sfrd-box}. Compared to CDM, star formation in the altDM models is suppressed at early times but eventually ``catches up'' and surpasses the CDM value in the later stages. This late-time starburst is particularly noticeable in the FDM model, where the SFR exceeds the CDM value by a factor of four times at $z=6$. The uniform UVB runs display a bump in star formation at $z\gtrsim 10$ regardless of the DM model employed. This is followed by a decline immediately after the UVB activation. Since the differences between DM models primarily arise from structure formation on small scales, they are more prominent in low-mass bins. In the high-mass bin, which is above the half-power mass of all three altDM models, the SFHs converge to the CDM results at $z \lesssim 10$.
    }
    \label{fig:sfrd-binned}
\end{figure*}

As shown in Figure~\ref{fig:rest_uvlf}, the UV luminosity functions predicted by the fiducial CDM \thesan model show no signs of flattening or bending at $M_{\rm 1500} \gtrsim -16$ at $z \geq 6$, which is consistent with the latest HFF results in \cite{Bouwens2022}. A cut-off in the UV luminosity function appears at $M_{\rm 1500} \sim -10$, but is likely caused by the limited mass resolution of stellar particles (see the similar feature in the stellar mass function in Figure~\ref{fig:smf}) and we do not explicitly model the physics that lead to a cut-off (e.g. dissociation of molecular hydrogen). 

The UV luminosity functions in altDM models show suppression at the faint end (above the resolution limit). The location of the suppression can be reasonably approximated by the half-power halo mass and the scaling relation described in Appendix~\ref{app:half-power}. The simulation results are consistent with existing observational constraints. JWST observations conducted in a lensed field will have the potential to reveal the faint-end suppression arising from altDM physics. However, the distinct signature of alternative DM in the UV luminosity function is less prominent than in the halo mass functions. Although the abundance of low-mass galaxies is suppressed in altDM models, the remaining ones are UV brighter. There are three primary reasons for this. (1) Delayed structure formation in altDM models leads to younger stellar populations and higher UV light-to-mass ratios. (2) Owing to the lack of progenitors below the half-power mass scale, haloes in altDM models, WDM and FDM in particular, predominantly assemble through roughly equal-mass (major), gas-rich mergers of haloes near the half-power mass. (3) The delay of reionization (relative to the CDM case under the same numerical configuration) results in a larger supply of cold, neutral, self-shielded gas available for future star formation once accreted into a galaxy. This effect is partly illustrated in Figure~\ref{fig:image-gas}). These factors collectively contribute to the accelerated formation of high-redshift galaxies in altDM models. Further evidence of this aspect will be shown in the following sections, discussing star formation efficiency and spatially resolved properties of galaxies. Similar phenomena have been found in many previous studies \citep[e.g.][]{Bose2016,Corasaniti2017,Dayal2017,Lovell2018,Lovell2019,Ni2019} with widely different theoretical approaches and galaxy formation models.

At $M_{\rm 1500} \lesssim -14$ at $z\geq 8$, all \thesanhr simulations fail to sample rare, UV bright sources and tend to underpredict the luminosity function compared to the \thesanone results and observational data. This is a systematic bias due to the limited volume of \thesanhr. Towards the faint end, this bias will eventually diminish and has little effect where the altDM signals appear. However, in addition to the bias, the cosmic variances here matter due to not only the small simulation volume but also the small lensed field in observations to push the detection limit down to $M_{\rm UV} \gtrsim -12$. We estimate this following the subsampling procedure described in Section~\ref{sec:hmf-smf} and the variance inflation factor calculated in Appendix~\ref{app:cosmic-var} (only the variance of halo number density is considered, so this is just a ``normalization'' variance). The effect of cosmic variance is illustrated with the dotted-dashed lines in the figure. The altDM signatures at the faint end could potentially be smeared at $z=6$ but are robust against cosmic variances at $z\gtrsim 8$. On the observational side, the typical field of view of plain field surveys is $10\,{\rm arcmin}$ (e.g. HUDF, \citealt{Beckwith2006}; XDF, \citealt{Illingworth2013}), which corresponds to a survey volume of $\sim (60\,{\rm cMpc})^{3}$ at $z \sim 7$ assuming a survey depth of $\Delta z = 1$. For lensed fields, the effective survey area will be much smaller on the source plane. For the HFF campaign~\citep{Lotz2017}, the typical survey area is $1\,{\rm arcmin}$ in the source plane at $z\sim 7$~\citep{Ishigaki2018}, which corresponds to a survey volume of $\sim (13\,{\rm cMpc})^{3}$. Neglecting other sources of observational uncertainties (e.g. completeness, lensing corrections), the cosmic variance will be less problematic compared to the small simulated volume.

In line with our findings for the stellar mass functions, the uniform UVB model suppresses galaxy abundance at the faint end following the activation of the UVB. By comparing the uniform UVB and the fiducial \thesanhr run, we effectively bracket the uncertainties arising from the patchy reionization process. In regions close to strong ionizing photon sources, the abundance of faint galaxies will be suppressed similarly to what we observe in the uniform UVB model. In contrast, for regions in voids, only local sources contribute to reionization, and the fiducial \thesanhr run represents this scenario. The uncertainties associated with the morphology of reionization are larger than the ``normalization'' uncertainties stemming from the variance of the matter density field (halo number count) as discussed in the paragraph above. Therefore, reliable modeling of the reionization process is essential for predictions of low-mass galaxy populations and quantitative comparisons with observational results.

\begin{figure*}
    \centering
    \includegraphics[width=1\linewidth]{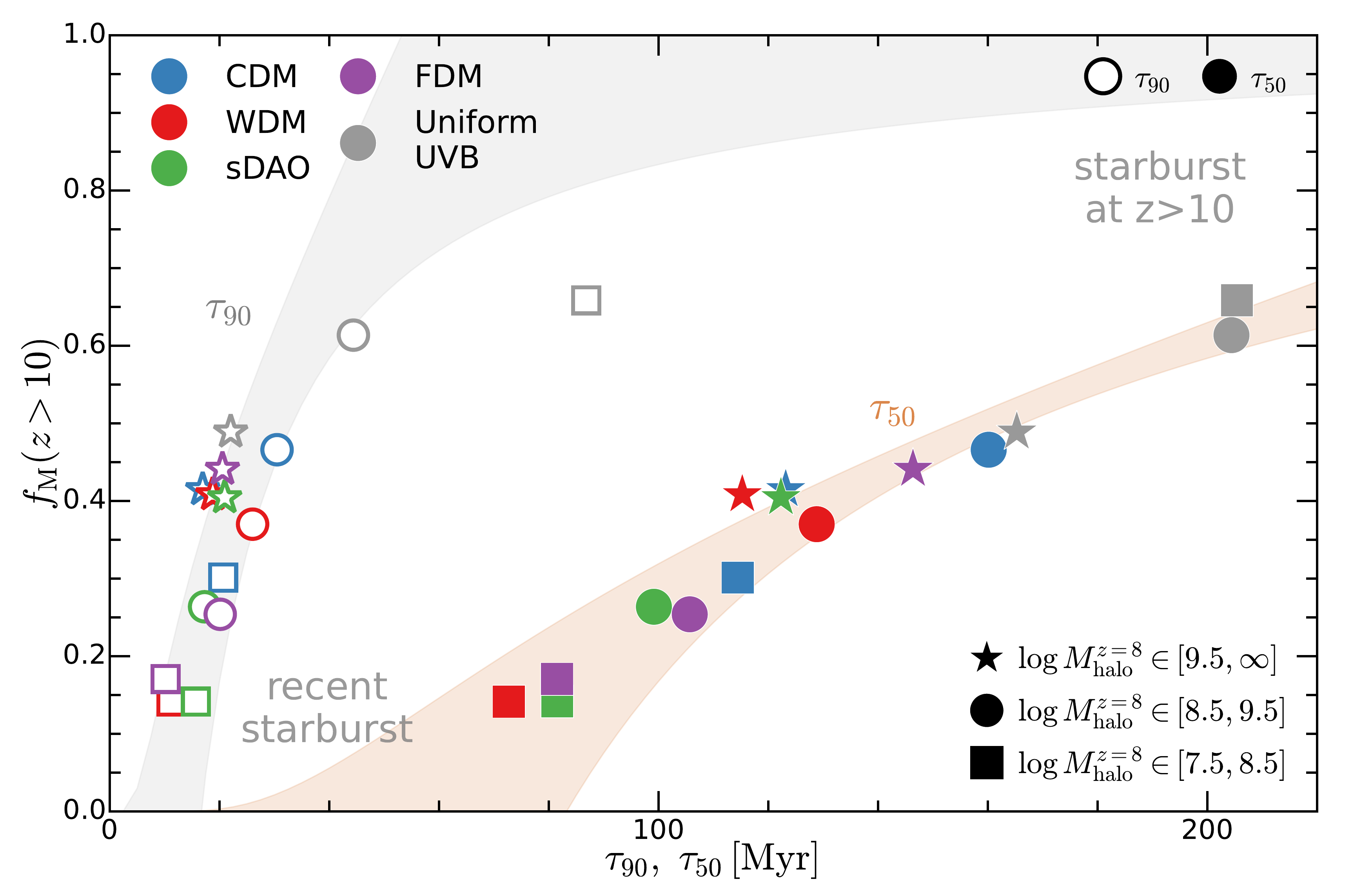}
    \caption{Fraction of stellar mass formed before $z=10$, $f_{\rm M}(z>10)$, versus the mass-weighted age of the stellar population ($\tau_{\rm x}$ is defined as the lookback time when ${\rm x}\%$ of the stellar mass is formed) for galaxies at $z=8$. Different DM or reionization models are distinguished by different colors, while haloes in different mass bins are distinguished by different markers (as labelled). In terms of the shaded band, the upper boundary represents the relationship obtained by assuming an exponentially rising SFH, while the lower boundary represents the results assuming a constant SFH. In general, points from RT simulations are above the lower boundary of this band, which indicates rising SFH. In relatively low-mass haloes, the stellar populations are younger and more dominated by the recent burst of star formation. In relatively massive haloes, the differences in stellar ages are smaller and the SFHs are close to or even steeper than exponentially rising. Conversely, the uniform UVB model (activated at $z\sim 10$) creates an artificial bump in the SFH at $z\gtrsim 10$ and a population of old stars.}
    \label{fig:sfh-z=8}
\end{figure*}

\section{Star formation \& Metal enrichment}
\label{sec:sfrd}

\subsection{Star formation history}

In Section~\ref{sec:visual} (see Figure~\ref{fig:sfrd-box}), we presented the history of the SFRD in the entire simulation volume. In Figure~\ref{fig:sfrd-binned}, we calculate the average SFH of haloes in three mass bins selected at $z=5$. The bin edges are chosen as $10^{8.5}\msun$ and $10^{9.5}\msun$. For reference, the half-power masses of the three altDM models studied lie between $10^{8.5}\msun$ and $10^{9.5}\msun$, as shown in Figure~\ref{fig:hmf}. In the high-mass bin, which is above the half-power masses of all three altDM models, there is a decrease in SFR at $z \gtrsim 10$ in altDM models, which can be attributed to the scarcity of low-mass progenitors at that time. However, at later times ($z\lesssim 10$), the average SFHs of haloes appear indistinguishable between the various DM models. In both the low-mass and intermediate-mass bins, we observe a systematic suppression of star formation in the early stages in altDM models. The strength of this suppression is positively correlated with the half-power mass, and the sDAO model, which has the largest half-power mass, demonstrates the most substantial suppression of SFR. However, at later times, starbursts occur in low-mass haloes in altDM models, allowing them to ``catch up'' with CDM and eventually surpass it. This observation aligns with our findings and discussions about the UV luminosity functions in Section~\ref{sec:uvlf}. In the FDM model, which features the steepest power spectrum damping, the average SFR in low-mass haloes can be approximately four times higher than that in CDM counterparts at $z=6$. The strength of this late-time starburst depends on the shape of the power spectrum damping, rather than simply scaling with the half-power mass. The enhanced star formation rates in altDM models could weaken the distinct signature of altDM in luminosity functions and global star formation/reionization history constraints \citep[e.g.][]{Lovell2019,Khimey2021}. The dashed lines in Figure~\ref{fig:sfrd-binned} show the SFH in the uniform UVB model. Similar to the global evolution of SFRD in this model, there is an increase (decrease) in SFR before (after) the activation of the UVB. This bump-to-suppression feature is amplified in the low-mass halo bin, as is the diversity of SFHs across different DM models. These artificial features in the SFH can contaminate the signature of altDM significantly, which emphasizes the importance of accurately modeling reionization at small scales.

To better illustrate the impact of altDM on properties of the stellar populations, we focus on galaxies at $z=8$ where the altDM models exhibit intense starbursts in low-mass haloes and clearly deviate from CDM, as demonstrated in Figure~\ref{fig:sfrd-binned}. We group them in three host halo mass bins $\log{(M^{z=8}_{\rm halo}/\msun)} \in [7.5,8.5],\,[8.5,9.5]$ or $[9.5,\infty]$. After stacking the SFH of the galaxies in each group, we determine three key metrics: (1) the fraction of stellar mass formed before $z=10$ denoted by $f{\rm M}$(z>10); (2) the lookback time that $50$\% of the stellar mass is formed (mass-weighted median age) represented by $\tau_{50}$; (3) the lookback time that $90$\% of the stellar mass is formed represented by $\tau_{90}$. In Figure~\ref{fig:sfh-z=8}, we show $f_{\rm M}(z>10)$ versus $\tau_{90}$ and $\tau_{50}$ in different models. For galaxies with $\log{(M^{z=8}_{\rm halo}/\msun)}<9.5$, altDM models generate younger stellar populations. For galaxies with $\log{(M^{z=8}_{\rm halo}/\msun)} \in [7.5,8.5]$, less than $20\%$ of the stellar mass are formed before $z=10$ and the median stellar age is about $80\Myr$ in the three altDM models, which is about $30\%$ lower than that of the CDM counterparts. These differences are weaker in relatively massive haloes with $\log{(M^{z=8}_{\rm halo}/\msun)}>9.5$, which is above the half-power mass of these models. The change in stellar age is systematic, with $\tau_{90}$ displaying similar trends as found for $\tau_{50}$. In the uniform UVB model, an artificial bump of star formation at $z\gtrsim 10$ causes $f_{\rm M}(z>10)$ and $\tau_{50}$ to increase up to approximately $0.6$ and $200\Myr$ respectively, regardless of the halo mass. The shaded regions in the figure show the relationship assuming a constant (lower boundary) and an exponentially-rising SFH (upper boundary). The fact that all the points (except for one in the uniform UVB model) lie between the relationship indicates galaxies in general have rising SFHs. The relatively massive galaxies tend to have more steeply-rising SFHs.

The SFH of galaxies will be constrained by JWST observations. Some attempts have been made on bright galaxies using SED fitting on legacy survey data \citep[e.g.][]{Laporte2021,Tacchella2022b,Whitler2023} or early JWST data \citep[e.g.][]{Tacchella2022,Furtak2023,Whitler2023z15}. They reveal the steeply rising SFHs of galaxies in the EoR that is qualitatively consistent with our simulation results. The exact age distribution of samples varies from studies and shows non-trivial dependence on the SFH priors \citep[e.g.][]{Tacchella2022b}.

\begin{figure}
    \centering
    \includegraphics[width=1\linewidth]{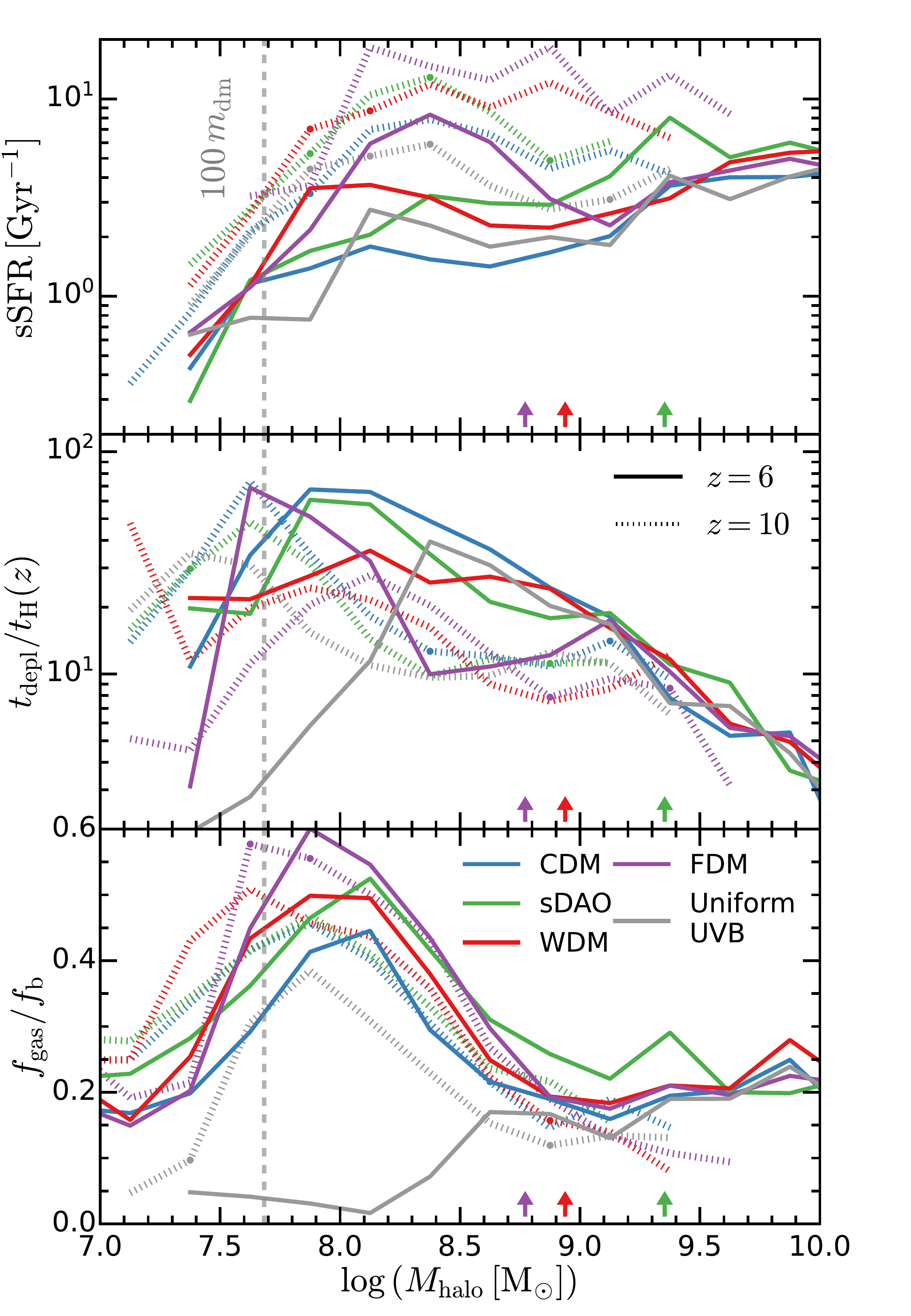}
    \caption{ {\itshape Top:} The specific star formation rate (sSFR) versus halo mass at $z=6$ (solid lines) and $z=10$ (dotted lines). In altDM models, the sSFRs are higher for halo masses below the half-power mass, $\sim 10^{9}\msun$, suggesting younger stellar populations. This enhancement is already noticeable at $z=10$. {\itshape Middle:} The gas depletion time $t_{\rm depl.}$ versus halo mass. The depletion time is defined as $M_{\rm gas}/{\rm SFR}$. In altDM models, the depletion times are systematically lower than in CDM below the half-power mass. This indicates that a larger fraction of gas in the halo is in a phase suitable for star formation. However, this trend is less evident at $z=10$. {\itshape Bottom:} The gas mass ratio $f_{\rm gas}$ versus halo mass. $f_{\rm gas}$ is normalized over the universal baryon fraction $f_{\rm b} \sim 0.16$. Gas is more abundant in altDM models likely due to more gas-rich major mergers. The uniform UVB model results in extremely small gas fractions in galaxies with $M_{\rm halo} < 10^{8.5}\msun$, leading to the rapid decline of depletion time observed in the middle panel.}
    \label{fig:tsf-vs-mhalo}
\end{figure}

\begin{figure*}
    \centering
    \includegraphics[width=0.485\linewidth]{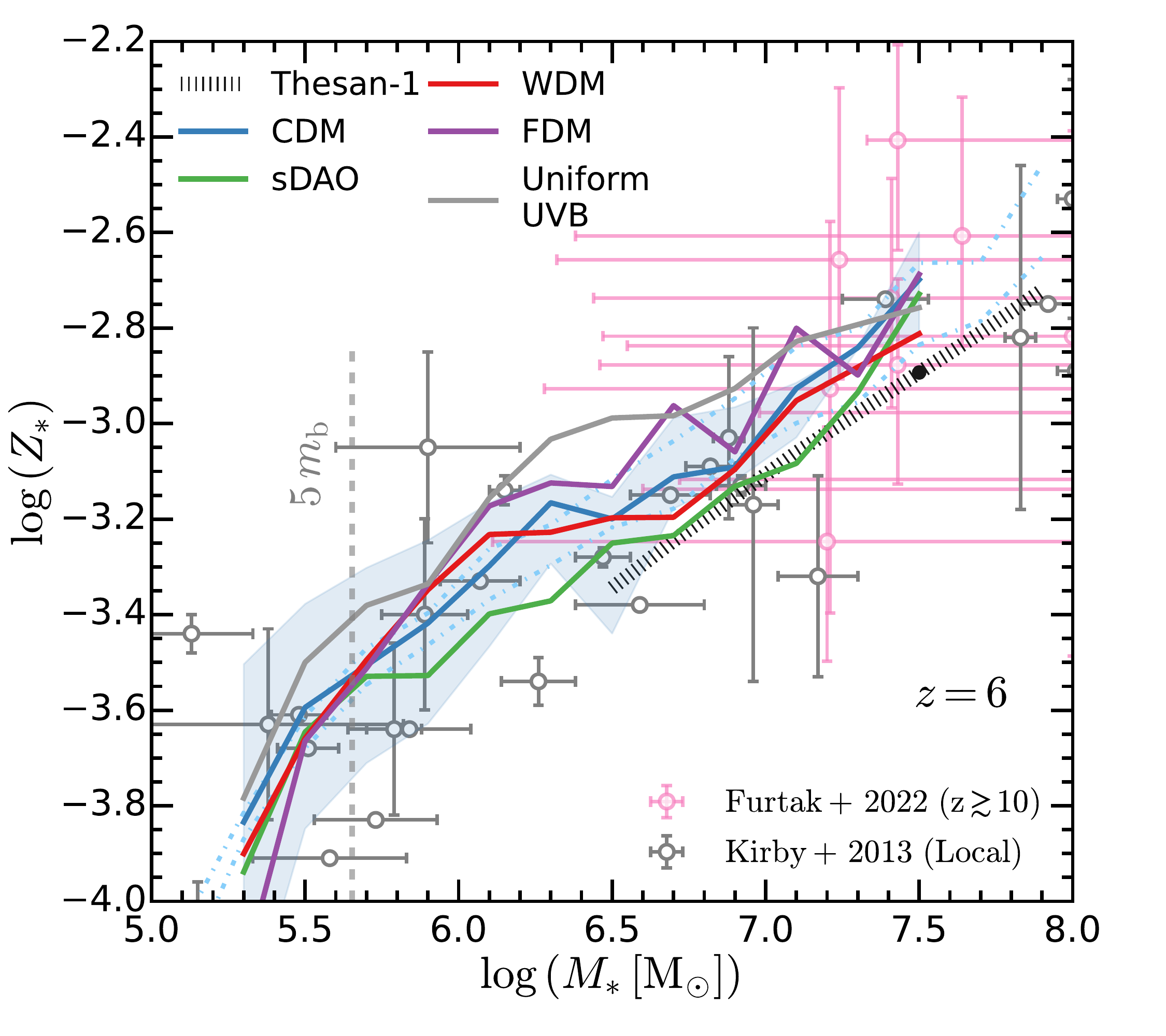}
    \includegraphics[width=0.485\linewidth]{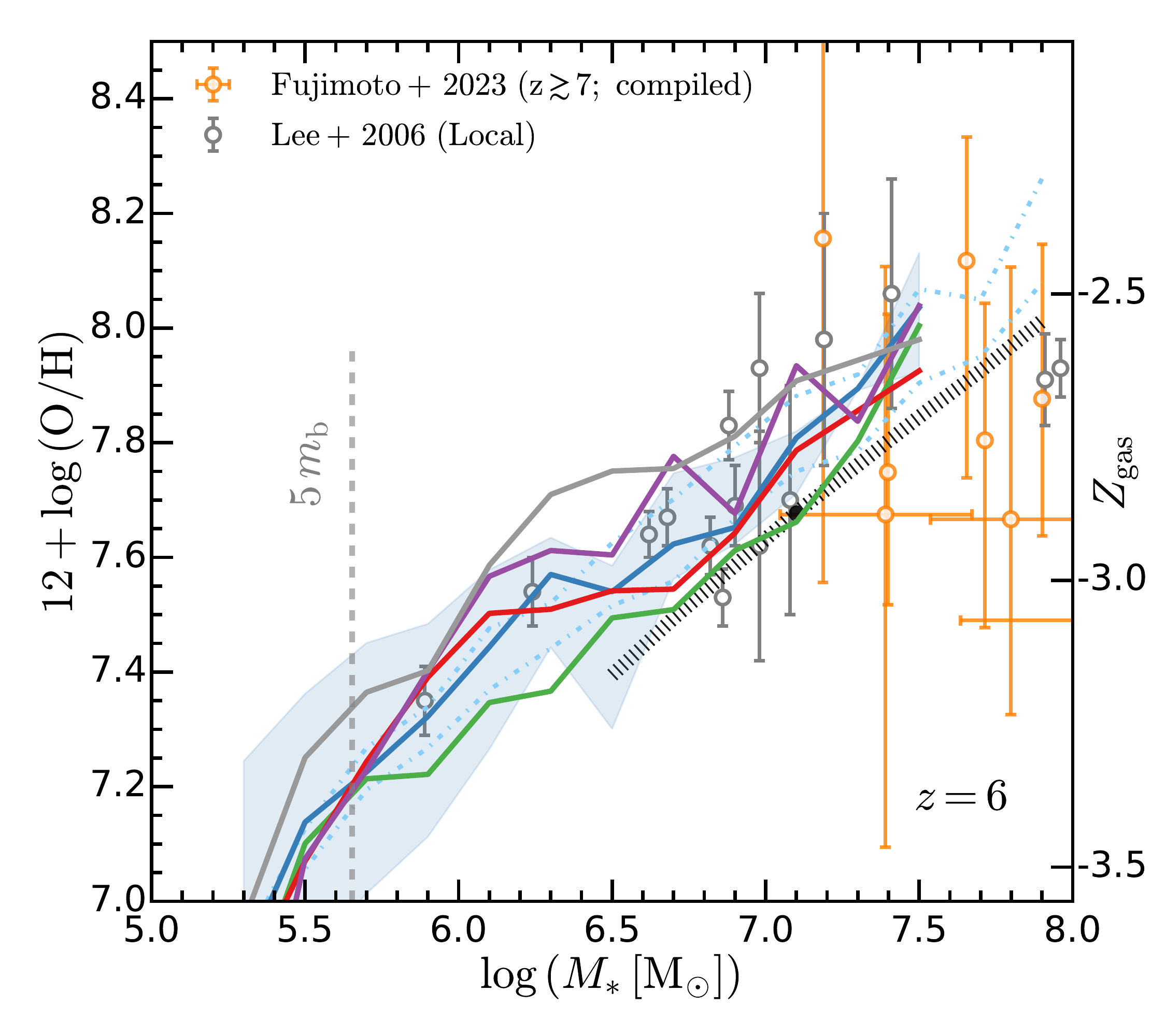}
    \caption{{\itshape Left:} Galaxy stellar metallicity versus stellar mass at $z=6$. The median metallicities are shown as solid lines. The $1\sigma$ dispersion is shown as the shaded region, for CDM only for clarity. \thesanone result is shown as the gray dotted line. The faint blue dash-dotted lines indicate cosmic variance. In addition, we show the early JWST measurements of $z\gtrsim 10$ galaxy candidates using SED fitting \citep{Furtak2023} and the measurements of the Local Group dwarfs compiled in \citet{Kirby2013}. {\itshape Right:} Galaxy gas metallicity versus stellar mass at $z=6$. The plotting style and label choices are the same as the left panel. For comparison, we show the recent JWST results from \citet{Fujimoto2023} as well as the local dwarf constraints from \citet{Lee2006}. The gas and stellar-phase metallicities resemble each other quite well. At $M_{\ast} \gtrsim 10^{7}\msun$, all the models give converged MZR. At $M_{\ast} \lesssim 10^{7}\msun$, galaxies in the sDAO model are less metal enriched by about $0.2\,{\rm dex}$ at mass scale $M_{\ast}\sim 10^{6 - 7}\msun$ while the ones in the uniform UVB model are more metal enriched by about the same level. Signatures are weaker for the WDM and FDM models since the recent burst of star formation can enrich galaxies to the same level as CDM or even more. The simulation results are in good agreement with the latest observational constraints of high redshift galaxies with JWST as well as the local dwarfs, within the scatter of the observed MZR.}
    \label{fig:mass-metallicity}
\end{figure*}

\subsection{Star formation efficiency}

In the top panel of Figure~\ref{fig:tsf-vs-mhalo}, we show the specific star formation rate (sSFR$\equiv {\rm SFR}/M_{\ast}$) as a function of halo mass. The SFR (instantaneous in gas cells) and $M_{\ast}$ are all measured using $R_{\rm max}$ as the aperture, following our definition of galaxy stellar mass in Section~\ref{sec:hmf-smf}. This choice is less affected by baryonic processes and their non-linear interplay with altDM physics. In general, altDM models predict higher sSFR in haloes below the half-power mass at both $z=6$ and $z=10$, which implies younger stellar populations. The enhancement in sSFR in low-mass haloes is more pronounced in models with steeper dampings of the linear matter power spectrum (i.e. FDM followed by WDM). As discussed in Section~\ref{sec:uvlf}, similar phenomena have been found in previous studies of high-redshift structure formation in altDM \citep[e.g.][]{Bose2016,Corasaniti2017,Lovell2018,Lovell2019}. But it is the first time that all relevant physics for cosmic reionization are included in a self-consistent fashion with a minimum level of fine-tuning. For example, in semi-analytical studies \citep[e.g.][]{Corasaniti2017}, the enhanced star formation efficiency in low-mass haloes is in fact a consequence of tuning the galaxy formation model in altDM to yield the correct timing of reionization, which is realized in our simulation out-of-the-box. In previous cosmological hydrodynamic simulations \citep[e.g.][]{Lovell2018,Ni2019}, where the uniform UVB model was employed, differences in the phase and morphology of reionization as part of the cause of the late-time starburst were not captured. In fact, \citet{Ni2019} found that the luminous fraction of haloes at $z\sim 6$ in FDM is lower than the CDM cases, which drives larger discrepancies in the UV luminosity function instead of closing the gap as we have found.

In the middle panel of Figure~\ref{fig:tsf-vs-mhalo}, we present the gas depletion time scale versus halo mass. The gas depletion time is defined as $t_{\rm depl} = M_{\rm gas}/{\rm SFR}$, where $M_{\rm gas}$ is defined as the total gas mass within $R_{\rm max}$. In low-mass galaxies, the depletion time is of order $10 - 100$ times the Hubble time $t_{\rm H}(z)$, comparable to the dynamical time of the haloes. It is consistent with the $\lesssim 1-10\%$ star formation efficiency per free-fall time from the local scaling relations~\citep[e.g.][]{Silk1997,KS1998}. In altDM models, the gas depletion times are shorter, consistent with their higher star formation efficiency, suggesting not only a larger gas supply but also a higher fraction of the gas in the star-forming phase. For the IllustrisTNG galaxy formation model employed, it means that the neutral gas density is closer to the star formation threshold. The trend is less obvious at $z=10$. In the uniform UVB model, the depletion time rapidly declines in low-mass haloes, but it is mainly due to the shortage of gas as will be shown in the bottom panel. 

In the bottom panel of Figure~\ref{fig:tsf-vs-mhalo}, we show the gas fraction (normalized using the universal baryon fraction, $f_{\rm b} \equiv \Omega_{\rm b}/\Omega_{\rm 0} \sim 0.16$) versus halo mass. $f_{\rm gas}/f_{\rm b}$ is much below unity since we only measure gas mass within $R_{\rm max}$ rather than the entire halo. The gas reservoir is more abundant in altDM models. The halo mass scale where they deviate from the CDM case is also consistent with the half-power mass. It supports the hypothesis that the haloes below the half-power mass in altDM are assembled mainly through gas-rich major mergers. In the uniform UVB model, the gas abundance is significantly suppressed at $M_{\rm halo} \lesssim 10^{8.5}\msun$ due to the strong ionizing photon bath, leading to the rapid decline of depletion time seen in the middle panel.

\subsection{Mass-metallicity relation}

The delayed star formation in altDM models will leave imprints on the metal enrichment of galaxies during or by the end of reionization. Direct observational constraints on galaxy stellar-phase metallicity can be inferred through SED fitting \citep[e.g.][]{Tacchella2022,CurtisLake2022,Furtak2023} and constraints on the gas-phase metallicity can be derived by emission line measurements \citep[e.g.][]{Curti2023,Fujimoto2023}. 

In the left panel of Figure~\ref{fig:mass-metallicity}, we show galaxy stellar metallicity versus stellar mass (known as the stellar-phase mass-metallicity relation, MZR) of simulated galaxies at $z=6$. Results at higher redshifts are similar. The sDAO model results in lower stellar metallicities than CDM at the 1$\sigma$ level for galaxies with $M_{\ast} \sim 10^{6 - 6.5}\msun$, due to the delayed structure formation. However, the WDM model yields an MZR comparable to that in CDM. The FDM model generates a similar MZR with slightly higher metal enrichment than CDM at $M_{\ast} \gtrsim 10^{6}\msun$. The MZRs in these models are affected by the late-time starbursts, which are not as pronounced in the case of sDAO. On the other hand, the uniform UVB model predicts a systematically higher MZR than the CDM run with RT. This discrepancy can be attributed to the artificial star formation that occurs at redshifts $z \gtrsim 10$ before the UVB is activated. 

For comparison, we show the early JWST results from \citet{Furtak2023} using SED fitting. The galaxies in this sample are in the redshift range $10\lesssim z \lesssim 15$ and are lensed behind the cluster SMACS J0723.3-7327 \citep{Atek2023} with the broadband photometric data from NIRCam and NIRISS fitted by {\sc BEAGLE}~\footnote{The code assumes $Z_{\odot} = 0.0154$ (different from the $Z_{\odot}=0.0127$ assumed by \thesan) and ${\rm [Fe/H]}\sim \log{(Z_{\ast}/Z_{\odot})}$.} \citep{BEAGLE}. The stellar metallicities of galaxies in simulations are in general consistent with these high-redshift sources. The low-mass galaxies at high redshift are quenched by the ionizing radiation background and their metal enrichment patterns will be almost frozen afterward. These galaxies are considered progenitors of the (ultra-)faint dwarf satellite galaxies observed in the Local Group (the Milky Way and M31, e.g. \citealt{Simon2019}). For comparison, we show the metallicity measurements of local dwarfs presented in \citet{Kirby2013}~\footnote{They assumed the Solar value $12 + ({\rm Fe/H})_{\odot} = 7.52$.}, which were derived from Fe abundances of individual stars. We directly convert $\rm [Fe/H]$ to $Z_{\ast}/Z_{\odot}$ (assuming $Z_{\odot}=0.0126$; \citealt{Asplund2004_solarabundance}). The MZRs of simulated galaxies in all the RT models are within the scatter of the observed local dwarfs at $M_{\ast} = 10^{5-7}\msun$. The uniform UVB model predicts higher MZR than the local dwarfs. This discrepancy will become even worse considering the potential late-time enrichment of dwarfs. 

At the relatively massive end ($M_{\ast} \gtrsim 10^{7}\msun$), the predictions of all models converge and fall within the scatter of the observed sample. The cosmic variance effects are estimated using the same method as in Figure~\ref{fig:hmf} and Figure~\ref{fig:rest_uvlf}. Meanwhile, there is a clear bias in the small-volume runs, with the \thesanone prediction being approximately $0.2$ dex below the \thesanhr results. This is due to the absence of massive bright sources in the small volume, which delays reionization and allows for stronger early-phase star formation and metal enrichment. It is worth noting that the observed MZR at $M_{\ast} \lesssim 10^{5}\msun$ flattens, which is in contrast to the prediction of cosmological simulations that suggest continuously decreasing metallicities in the low-mass end \citep[e.g.][]{Wheeler2019}. However, the baryonic mass resolution in our simulations is not sufficient to investigate this phenomenon. In future studies, it would be intriguing to explore how patchy reionization or alternative dark matter models might impact the metal enrichment of these low-mass galaxies.

In the right panel of Figure~\ref{fig:mass-metallicity}, we show the gas-phase MZR of galaxies in simulations. The hydrogen and oxygen abundance are directly taken from gas cells in simulations. The gas-phase MZRs in simulations closely resemble the stellar-phase ones. The Type-{\small II} supernovae rates and the continuous mass-loss rate through Asymptotic Giant Branch (AGB) stellar winds drop significantly by $\sim 100\Myr$ \citep[e.g.][]{Portinari1998,Leitherer1999,Vogelsberger2013}, both being the dominant sources of metal enrichment. Therefore, the metal recycling process happens at a relatively small time scale compared to the lifetime of these galaxies. Since these galaxies are predominantly influenced by the recent period of star formation, it is not surprising that the stellar and gas-phase metallicities show great synergy. The simulation results are compared to the JWST spectroscopically-confirmed sources in \citet{Fujimoto2023} and the Local Group dwarfs in \citet{Lee2006}. The sources in \citet{Fujimoto2023} are measured using O[{\small III}] and H$\beta$ emission lines as tracers. The simulated massive galaxies exhibit slightly higher gas metallicities compared to the JWST sources but within the scatter. Overall, the influence of altDM models is not statistically significant on the MZR plane. A major reason for this is that the late-time starbursts in altDM models wash out the altDM signatures in metal enrichment levels. The precision and statistics of current observations are not sufficient to differentiate between altDM models.

\begin{figure}
    \centering
    \includegraphics[width=1\linewidth]{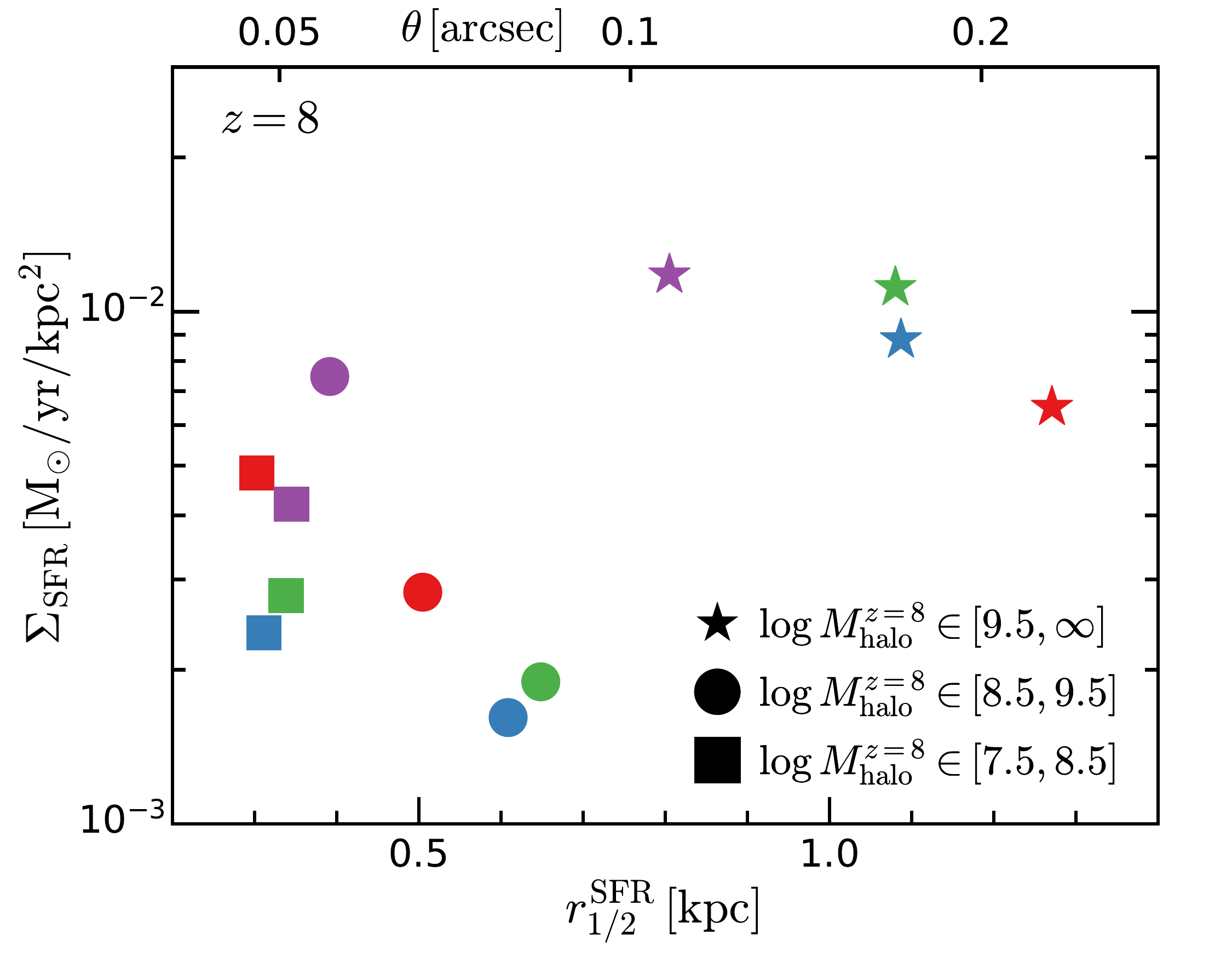}
    \includegraphics[width=1\linewidth]{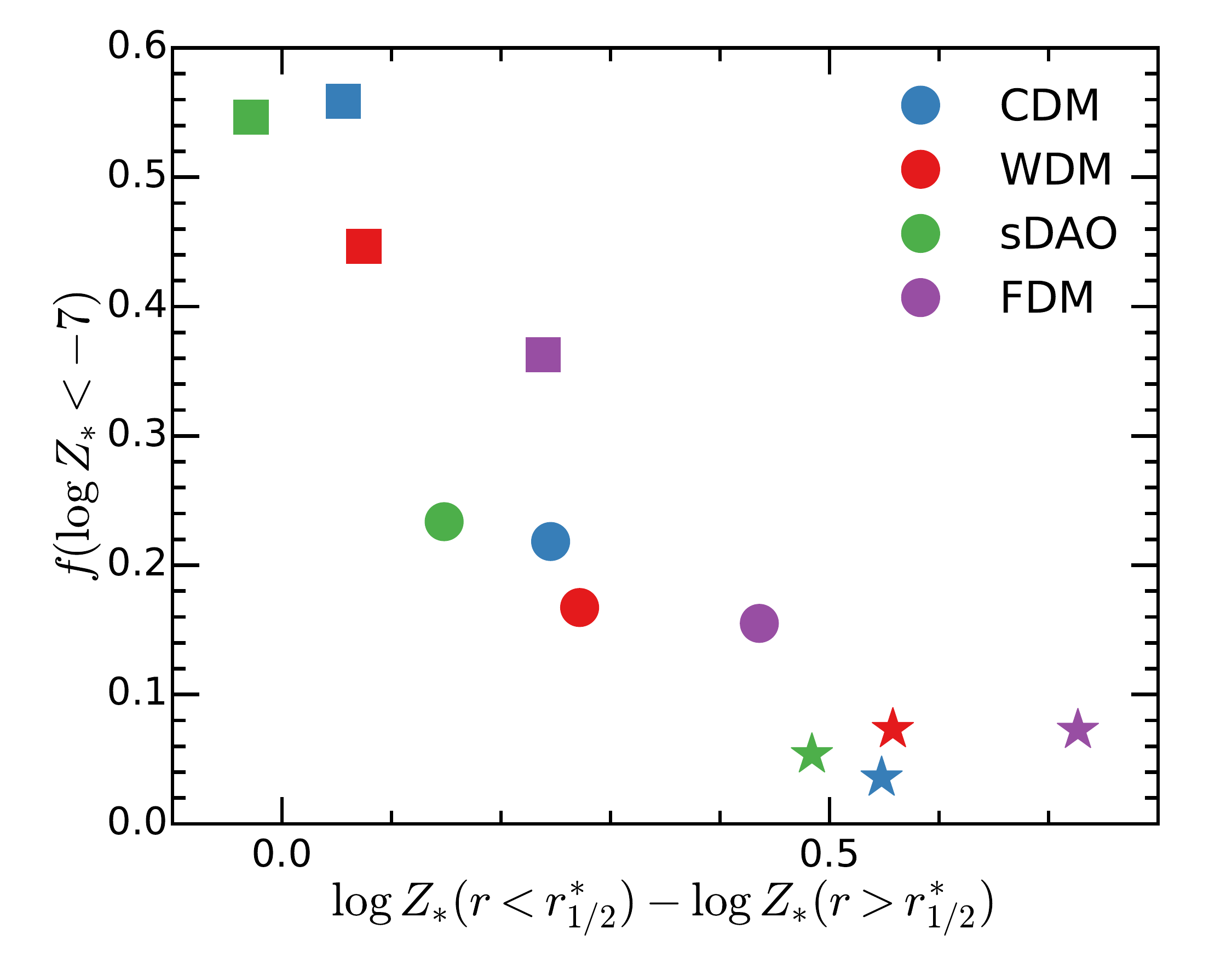}
    \caption{{\itshape Top}: Surface density of SFR versus the half-SFR radius for galaxies at $z=8$ stacked in three halo mass bins. The half-SFR radius is a proxy for the half-light radius in rest-frame UV which traces recent star formation. We combine the SFR calculated using the instantaneous SFR in gas cells and the one using masses of young stars with age $\lesssim 100\Myr$. The FDM model results in smaller galaxy sizes and enhanced SFR surface density. The signature is less prominent in the WDM and sDAO models. {\itshape Bottom}: Fraction of stellar mass with metallicity below $10^{-6}$ versus the metallicity gradient of the same groups of galaxies in the top panel. The metallicity gradient is represented by the difference in the mean metallicity measured within and outside $r^{\ast}_{1/2}$. Galaxies in the WDM and FDM models have larger metallicity gradients. In low-mass haloes, the fractions of low-metallicity stars are appreciably smaller in WDM and FDM. These findings are consistent with the late-time starbursts, which predominantly occur in the central regions of galaxies.
    }
    \label{fig:spatially-resolved}
\end{figure}

\subsection{Spatially-resolved properties}

The late-time starburst found in altDM models can influence the stellar distribution within galaxies and the relative contribution of in-situ and ex-situ star formation. In the following, we investigate the spatially resolved properties of galaxies. To enhance the statistics of stellar particles and gas cells, we stack galaxies at $z=8$ in three halo mass bins, $\log{(M^{z=8}_{\rm halo}/\msun)} \in [7.5,8.5],\,[8.5,9.5]\,[9.5, \infty]$. 

To quantify galaxy sizes, we measure the half-SFR radius ($r^{\rm SFR}_{1/2}$), which encloses half of the total SFR within the halo. We combine two proxies for SFR, the instantaneous SFR in gas cells and the archaeological SFR from young stellar particles (with age $\lesssim 100\Myr$). The half-SFR radius offers a better analogy to the half-light radius observed in rest-frame UV, although projection effects could still cause some order-unity differences. In the top panel of Figure~\ref{fig:spatially-resolved}, we show the SFR surface density (calculated within $r^{\rm SFR}{1/2}$) as a function of $r^{\rm SFR}{1/2}$. The $r^{\rm SFR}_{1/2}$ is translated to angular distance on the sky using the angular diameter distance based on the assumed cosmology. In the FDM model where the late-time starburst is most prominent, we find more compact galaxy sizes in host haloes with $\log{(M_{\rm halo}/M_{\odot})} \gtrsim 8.5$ compared to their CDM counterparts. In addition, we find enhanced SFR surface densities across all three halo mass bins, including the one above the half-power mass. This increased compactness is likely a result of rapid ongoing star formation in the central regions of galaxies fueled by a series of gas-rich major mergers, as well as a reduced mass of ex-situ stars formed in progenitors. We find similar signatures in the WDM model for relatively low-mass haloes ($\log{(M_{\rm halo}/M_{\odot})} \lesssim 8.5$). However, the sDAO model exhibits less apparent signatures, as it does not show comparably strong late-time starbursts at $z=8$ (see Figure~\ref{fig:sfrd-binned}).

Compact galaxy sizes (roughly $100\,{\rm pc}$ to $1\kpc$) are found for the low-mass ($M_{\ast}\lesssim 10^{7}\msun$) galaxies in our simulations, regardless of the DM models. These results are consistent with extrapolations (in terms of stellar mass and redshift) of observational findings \citep[e.g.][]{vanderWel2014,Shibuya2015} and theoretical expectations \citep[e.g.][]{Mo1998}. Galaxy size is an essential factor in forward modeling lense-magnified sources and determining the luminosity function on the image plane \citep[e.g.][]{Bouwens2017,Kawamata2018,Ishigaki2018,Bouwens2022-size}. The upcoming JWST observations will offer more robust morphological constraints on these compact, low-luminosity galaxies. Attempts have been made for the relatively luminous sources \citep[e.g.][]{Ono2022, Robertson2022, Yang2022, Tacchella2023}. These early JWST sources at $M_{\rm UV} \lesssim -18$ are all compact with effective radii between $100\,{\rm pc}$ and $1\kpc$. The face values are consistent with our results, but the extrapolation to faint luminosities and potential selection biases are still uncertain.

In the bottom panel of Figure~\ref{fig:spatially-resolved}, we show the fraction of low-metallicity stars ($\log{Z_{\ast}}< -7$) versus the stellar metallicity gradient at the outskirt of galaxies. The metallicity gradient is estimated by calculating the difference between the mass-weighted stellar metallicity measured within and outside the stellar-half-mass radius $r^{\ast}_{1/2}$. In all three halo mass bins, galaxies in the FDM model exhibit larger stellar metallicity gradients. Similar to what we found for galaxy sizes, the difference is evident above the half-power mass, where the halo mass function has already converged to the CDM results. The metallicity gradient is also driven by centrally dominated late-time star formation as well as lowered metallicities of the ex-situ population of stars at the outskirt of galaxies. Similar trends are found for gas-phase metallicity gradients as well. These signatures are less noticeable in the WDM model and are absent in the sDAO model, where the late-time starburst is not comparably strong at $z=8$. Furthermore, in low-mass haloes with $\log{(M_{\rm halo}/\msun)} \in [7.5,8.5]$, the WDM and FDM models show a smaller fraction of extremely low-metallicity stars compared to CDM, which are inherited from the earliest generation of star formation.

\section{Can we disentangle DM physics with the uncertainties in the baryonic sector?}
\label{sec:discuss}

In addition to the altDM physics explored in this paper, several other astrophysical factors could also influence the faint-end UV luminosity functions as well as properties of low-mass galaxies in the EoR. The leading three are discussed in the following.

(1) {\itshape Atomic cooling limit.} The low-temperature molecular hydrogen (which is responsible for gas cooling at low temperatures, $T\lesssim 10^{4}\,{\rm K}$) can be dissociated with a moderate strength of LW radiation background sourced by Pop{\small III} stars \citep[e.g.][]{Haiman1997,Haiman2000,Machacek2001}. Therefore, Pop{\small II} star formation can only take place in relatively massive haloes ($\gtrsim 10^{6}\msun$) that can host a reservoir of self-shielded, molecular gas or with atomic cooling channels enabled \citep[e.g.][]{Wise2007,Greif2008, Oshea2008}. Flattening of the faint-end luminosity functions were predicted at $M_{\rm UV} \gtrsim -12$ in simulations with explicit models of Pop{\small III} star formation \& feedback and LW radiation background~\citep[e.g.][]{Wise2014,Oshea2015,Ocvirk2016}. In \thesanhr, the molecular phase of gas and low-temperature cooling are not directly modeled but are partially captured by the effective equation of state of the ISM gas \citep{Springel2003}. As a result, the birth halo masses of stellar particles in our simulations are all above $\sim 10^{7.5}\msun$. The mass limit is empirically consistent with the atomic cooling limit ($T_{\rm vir}\sim 10^{4}\,{\rm K}$) but larger than the limit found in e.g. \citet{Wise2007} due to the lack of Pop{\small III} star physics.
    
(2) {\itshape Feedback from star formation.} This includes stellar (pre-supernovae) and supernovae feedback that can drive galactic-scale winds as well as radiative feedback from young massive stars \citep[e.g.][]{Gnedin2000,Wise2009,Hopkins2013,Vogelsberger2013}. Feedback is important in regulating star formation within the halo, leading to systematically lower star formation efficiency in low-mass haloes \citep[e.g.][]{Behroozi2013}. Negative feedback changes the kinematics and the phase of the gas, which limits the effectiveness of gas cooling and gradually boils away the star-forming gas reservoir \citep[e.g.][]{Efstathiou1992,Thoul1996,Dijkstra2004,Shapiro2004}. Radiative feedback such as photoionization and photoheating can have effects on large scales, reducing the neutral gas supply of galaxies in the environment and suppressing star formation activities at later times \citep[e.g.][]{Jeon2014,Jeon2015}. These mechanisms all tend to reduce the stellar mass of low-mass haloes or equivalently the number density of galaxies of fixed stellar mass. On the other hand, as galaxies grow more massive, the feedback is eventually dominated by local sources and star formation becomes self-regulating \citep{Wise2008, Schaye2010, Ostriker2010, Hopkins2014, Rosdahl2015, Hopkins2023}, removing the signatures of altDM models at late stages of galaxy evolution.
    
(3) {\itshape Morphology of reionization.} The phase and morphology of reionization have a strong influence on the abundance of faint galaxies, as emphasized in the previous \thesanhr study \citep{Borrow2022}. In volumes embedded in the ionizing bubble created by massive bright galaxies, ionizing radiation can deplete the reservoir of cold, neutral gas in low-mass galaxies, thereby suppressing star formation. For example, in the uniform UVB model we tested, the faint-end UV luminosity function is elevated (suppressed) before (after) the activation of the UVB, at similar luminosities where altDM signatures appear. We found related signatures in the SFRD evolution and stellar mass functions discussed in previous sections. Conversely, if the volume is situated in a void of ionizing sources (as effectively occurs in the fiducial \thesanhr runs), the phase of reionization will be delayed, creating a more favorable environment for star formation in low-mass galaxies. This introduces environment-dependent bias and an additional source of uncertainty in the galaxy-halo connection, resulting larger cosmic variance of the UV luminosity functions than the halo mass function.

The physical processes outlined above can generate similar suppression features in the faint end of galaxy UV luminosity function. However, they all impose {\itshape negative} feedback on star formation in a {\itshape coherent and systematic} manner. For instance, a strong, uniform bath of ionizing radiation from bright sources in the environment will consistently advance the phase of reionization and suppress star formation at all stages of reionization. Similarly, a model featuring stronger stellar winds or supernovae or radiative feedback from local sources tends to push gas out of the star-forming phase and suppress star formation in galaxies. For both cases, the negative feedback on star formation is independent of the global phase of halo assembly. 

However, as we found for altDM models, a late-time starburst takes place in low-mass galaxies and allows altDM universes to rapidly ``catch up'' the global phase of reionization. It is effectively a {\itshape positive} feedback that is only associated with low-mass galaxies at late times of the EoR. Importantly, this phenomenon is rooted in the hierarchical assembly pattern of DM haloes rather than merely altering the connection between gas/star formation and haloes. Meanwhile, it is linked to various changes in galaxy properties we observed, including higher star formation efficiency (resulting in younger stellar populations), higher SFR surface density, more compact galaxy sizes, and stronger metallicity gradients at the outskirts of galaxies. The signatures of altDM could be disentangled with uncertainties arising from astrophysical processes through joint analysis of luminosity function measurements as well as constraints on galaxy SFH and morphology. 

Independent constraints on reionization will also be beneficial. The global signal of the ionized phase from e.g. CMB optical depth, Lyman-$\alpha$ emitters at high redshift can help constrain the sub-grid model parameters of the reionization model. In addition, for luminosity function measurements in gravitationally lensed fields, it is important to constrain the local phase of reionization. It will be particularly helpful to reduce the uncertainties from the morphology of reionization. Such measurements can be achieved through Lyman-$\alpha$ forest tomography \citep[e.g.][]{Delubac2015,Yang2020,Bosman2022} and the intensity mapping of the 21 cm spin-flip transition of the hydrogen atom \citep{Furlanetto2006, Mellema2006, Parsons2010, Pritchard2012}. 

\section{Conclusions}
\label{sec:conclusion}

In this paper, we utilized \thesanhr, a suite of high-resolution, cosmological, hydrodynamic simulations of galaxies in the EoR, to explore the impact of alternative DM models that suppress the linear matter power spectrum. The inclusion of the radiative transfer method alongside the galaxy formation model and the high numeric resolution featured in these simulations provide a solid foundation for us to study various galaxy properties as well as the morphology of reionization in altDM models. In this paper, we focus on the properties of galaxies that can be revealed by the JWST. Our major findings can be summarized as follows.

\begin{itemize}
    \item {\itshape Galaxy abundance:} In altDM models, the imprint of the suppression in the linear matter power spectrum is found in the halo mass functions, the stellar mass functions, and the UV luminosity functions at the low-mass/faint end. The mass/luminosity scale of the suppression correlates well with the half-power wavenumber $k_{\rm 1/2}$ (see Appendix~\ref{app:half-power} for details). In addition, the shape of the suppression in galaxy abundance reflects the shape of the damping wing of the matter power spectrum. For example, a steeper damping of the power spectrum in FDM compared to WDM results in a steeper suppression in the mass functions as well. The residual acoustic oscillation features result in continuously rising mass functions in the low-mass end. JWST observations of lense-magnified fields are promising in detecting the suppression features due to altDM models, but one needs to be aware of various astrophysical effects influencing low-mass galaxy abundance.
    
    \item {\itshape Complexities from reionization modeling:} In terms of the abundance of faint galaxies, one important astrophysical uncertainty is the morphology of the reionization process. In simulations using a uniform UVB model, a strong bath of ionizing photons contributed by massive bright non-local sources is applied to low-mass haloes. As a result, the abundance of low-mass, faint galaxies at $M_{\ast} \lesssim 10^{6}\msun$ or $M_{\rm 1500}\gtrsim -14$ are suppressed. The location of this suppression is similar to where altDM signatures show up. Therefore, it is important to correctly model the morphology of reionization to make robust predictions for the faint-end luminosity functions and use that to constrain the nature of DM.
    
    \item {\itshape Complexities from non-linear effects:} In all altDM models, in particular the ones with steep power spectrum dampings, we find positive feedback on the late-time star formation in galaxies with host halo mass below the half-power mass. This manifests as brighter UV luminosities, higher sSFR, and gas abundance as well as shorter gas depletion time in haloes below the half-power mass scale. The late-time starbursts allow the SFRD and ionizing photon production rates in altDM to ``catch up'' with or even surpass the CDM value at lower redshifts. Notably, the volume-weighted ionized fraction in the FDM model is accelerated to exceed the CDM counterpart at $z \sim 6$.
    
    \item {\itshape Metal enrichment}: The delayed star formation in altDM models can lead to suppression of metal abundances in low-mass galaxies until the late-time starbursts take over. For the sDAO model, the signal is at the $1\sigma$ level at the mass range $M_{\ast} \sim 10^{6-6.5}$ at $z=6$. The WDM and FDM models on the other hand experience a boost in late-time star formation and metal enrichment, and can thus ``catch up'' with CDM or even exceed it. However, the differences are still encoded in the spatial distribution of metals in galaxies. In altDM models, especially the FDM model, stellar/gas metallicities show stronger gradients at the outskirts of galaxies, and the population of ``old'' stars with extremely-low metallicities is suppressed. 
    
    \item {\itshape Galaxy sizes}: In altDM models, we find more compact galaxy sizes in low-mass haloes below the half-power mass. This is associated with enhanced SFR surface density. The differences are driven by the late-time starburst, which takes place in the central regions of galaxies, as a consequence of more frequent gas-rich major mergers in altDM models.

    \item {\itshape Ways to disentangle DM physics and astrophysical uncertainties:} Many astrophysical processes could generate suppression features in the faint-end luminosity function, which can contaminate the altDM signatures. Two examples are a strong ionizing photon background and stronger stellar and supernovae feedback. Most of these processes will cast negative feedback on star formation in a coherent and systematic fashion. However, for altDM models, we find positive feedback on star formation that is only associated with low-mass galaxies at late times of the EoR. It is a unique feature of altDM physics that delays early structure formation and is linked to various changes in galaxy SFH and morphology. A joint analysis of the luminosity functions and galaxy properties could help disentangle these effects.

\end{itemize}

In this study, we investigated the properties of faint galaxies during the EoR with a focus on the influence of altDM models. Our findings corroborate several previous assertions, such as the reduced abundance of faint galaxies and the promise of detecting this with JWST observations. By combining the radiation-hydrodynamic solver and the IllustrisTNG galaxy formation model, we can study the effects of reionization on low-mass galaxies, as well as their reciprocal influence on reionization, within altDM models. We identify enhanced late-time star formation in altDM models, consistent with earlier studies, which arises from both different patterns of hierarchical assembly and delayed reionization process in altDM. This can introduce non-linear effects on the reionization history at small scales. We also draw attention to the complexities inherent in both reionization and galaxy formation models and propose ways to disentangle them with altDM physics.

Looking ahead, it is crucial to further explore the impact of baryonic physics in a more controlled setting. We plan to conduct zoom-in RHD simulations with multi-phase ISM physics, allowing us to investigate the impact of diverse star formation and feedback prescriptions on low-mass galaxies, providing valuable insights into the interplay between the ``luminous'' and DM physics. In addition, simulations with larger and more representative volumes will allow us to make quantitative predictions for the properties of the IGM. This will enable us to examine metrics such as the HI clumping factor and make predictions for the Lyman-$\alpha$ forest and 21 cm line intensity mapping.

\section*{Acknowledgements}
The computations in this paper were run on the Engaging cluster at Massachusetts Institute of Technology (MIT) as well as the Faculty of Arts and Sciences Research Computing (FASRC) Cannon cluster at Harvard University. MV acknowledges support through the National Aeronautics and Space Administration (NASA) Astrophysics Theory Program (ATP) 19-ATP19-0019, 19-ATP19-0020, 19-ATP19-0167, and NSF grants AST-1814053, AST-1814259, AST-1909831, AST-2007355 and AST-2107724. 
EG acknowledges support from the Canon Foundation Europe and the University of Osaka during part of this research.
JZ acknowledges support from a Project Grant from the Icelandic Research Fund (grant number 206930). \\

\noindent Software citations:
\begin{itemize}
    \item {\sc Numpy}:    \citet{Harris2020}
    \item {\sc Scipy}:    \citet{Virtanen2020}
    \item {\sc Astropy}:  \citet{Astropy2013,Astropy2018,Astropy2022}
    \item {\sc Matplotlib}: \citet{Hunter2007}
    \item {\sc Swiftsimio}: \citet{Borrow2020, Borrow2021}
    \item {\sc Arepo-rt}: \citet{Springel2010, Kannan2019, Weinberger2020}
\end{itemize}

\section*{Data Availability}
All \thesan and \thesanhr simulation data will be made publicly available in the near future. Data will be distributed via \href{project website}{www.thesan-project.com}. The post-processed data and analysis scripts are stored on the Engaging cluster at MIT. Before the public data release, the data underlying this article can be shared on reasonable request to the corresponding author.







\appendix

\section{Large-scale structures and IGM in the uniform UVB runs}
\label{app:visual}

\begin{figure}
    \centering
    \includegraphics[width = 1\linewidth]{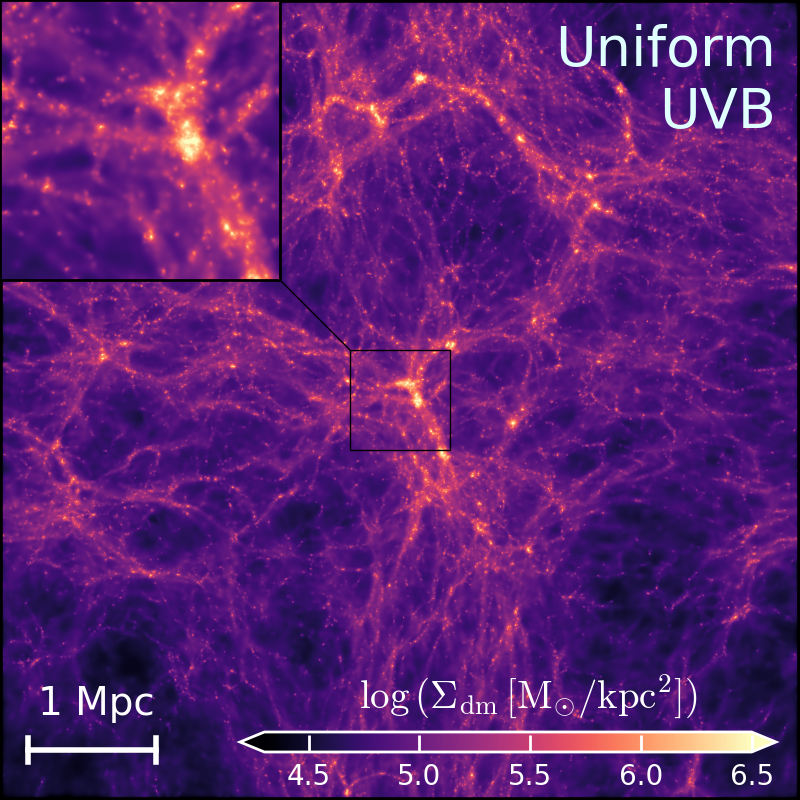}
    \includegraphics[width = 1\linewidth]{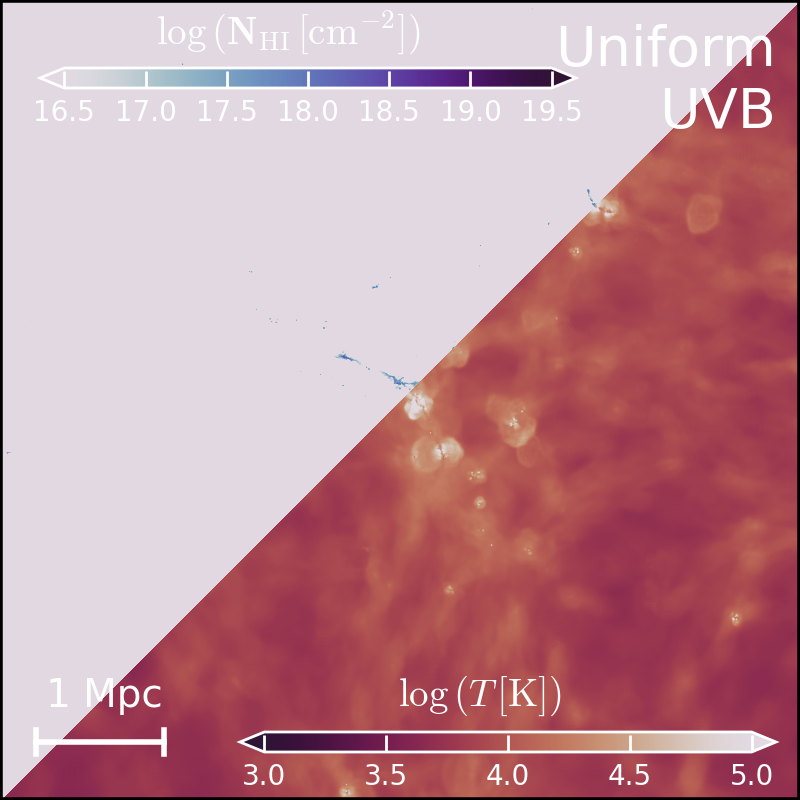}
    \caption{{\itshape Top}: DM surface density in the CDM--uniform UVB run at $z=6$. The DM distribution on large scales is nearly identical to the fiducial CDM run using \thesan physics. {\itshape Bottom}: Neutral hydrogen column density and gas temperature distribution at $z=6$. The IGM is almost completely ionized after the uniform UVB is activated, with the exception of some dense, self-shielded gas in the most massive halo and the surrounding filaments. The IGM temperature has been elevated to $\gtrsim 10^{4}\,{\rm K}$ by the strong ionizing radiation bath.}
    \label{app-fig:uvb}
\end{figure}

In the top panel of Figure~\ref{app-fig:uvb}, we show the DM surface density map in the uniform UVB model for CDM at $z=6$. The DM distribution is identical to that in the fiducial \thesanhr simulations using RT, as presented in Figure~\ref{fig:image-dm}. However, the ionization state and thermal properties of the IGM can be significantly altered by the reionization model. In the bottom panel of Figure~\ref{app-fig:uvb}, we show the neutral hydrogen column density and gas temperature distribution of the IGM at $z=6$ in the uniform UVB model. Following the activation of the UVB at $z\sim 10$, the IGM becomes almost completely ionized, except for a few dense, self-shielded gas clumps in massive haloes and their surrounding filaments. Meanwhile, the IGM temperature rises to $\gtrsim 10^{4}\,{\rm K}$ due to the intense ionizing radiation background. These changes to the IGM thermal properties influence the supply of cold neutral gas, which fuels late-time star formation in galaxies, and results in the suppressed SFRD and faint-end luminosity functions in the uniform UVB model at $z\lesssim 10$.

\section{The characteristic mass/luminosity of altDM signatures}
\label{app:half-power}

As shown in Section~\ref{sec:hmf-smf}, the low-mass end of the halo/stellar mass function is directly affected by the suppression of the small-scale power spectrum in altDM models. Here we provide simple estimates of the characteristic halo/stellar mass of the suppression feature.

In \citet{Bohr2021}, the Press-Schechter formalism \citep[including its variants;][]{Press1974, Bond1991, Sheth1999, Sheth2001} with a smooth {\itshape k}-space window function \citep{Leo2018} was shown to predict the halo mass function in altDM models with good accuracy. The corresponding halo mass scale of a given wavenumber in the power spectrum is
\begin{align}
    M_{\rm halo}(k) &= \dfrac{4\pi \bar{\rho}_{\rm m}}{3}\,\dfrac{c_{\rm w}}{k^{3}}\nonumber \\
    &= 9.3\times 10^{9} \msun \left( \dfrac{\Omega_{\rm m}}{0.3}\right)^{2}\, \left( \dfrac{h}{0.7}\right)^{2} \left( \dfrac{k}{10\,h\Mpc^{-1}}\right)^{-3},
    \label{eq:mcut}
\end{align}
where $c_{\rm w}=3.79$ is a constant for translation of comoving scale and wavenumber as suggested in \citet{Leo2018} ($c_{\rm w}=\pi$ is used in some literature, but the difference is small for the purpose of this study). Plugging in the half-power wavenumber $k_{1/2}$ of altDM models gives a reference for the halo mass where the suppression will occur (half-power mass, as quoted in the main text). For stellar mass functions, we can estimate the mass scale corresponding to a wavenumber as $M_{\ast}(k) = f_{\ast} \,M_{\rm halo}(k) \equiv \bar{\epsilon}_{\ast} \,f_{\rm b} \,M_{\rm halo}(k)$, where $f_{\ast}$ is the stellar-to-halo-mass ratio, $f_{\rm b}$ is the universal baryon fraction, $\bar{\epsilon}_{\ast}$ is the averaged star-forming efficiency of gas accreted in haloes (in the growth history of low-mass haloes). At the halo mass scale of around $10^{9}\msun$, $f_{\ast}$ is estimated to be around $10^{-3}$ by extrapolating the results of abundance matching and empirical modeling \citep[e.g.][]{Behroozi2013, Behroozi2019}. In addition, in empirical models specifically constrained for high-redshift galaxies \citep[e.g.][]{Tacchella2013, Mason2015, Tacchella2018}, a consistent value of $\bar{\epsilon}_{\ast}\,f_{\rm b} \sim 10^{-3}$ is found for low-mass haloes. We will use these formulae to estimate the characteristic halo (stellar) mass scale where we expect differences between DM models to show up. 

To further relate this to the suppression of the rest-frame UV luminosity functions, we adopt the scaling relation in \citet{Dayal2015}
\begin{equation}
    \log{M_{\ast}} = \beta M_{\rm UV} + \gamma, 
    \label{eq:mstar-muv}
\end{equation}
where $\beta = -0.38$, $\gamma = \gamma_{0} - 0.1z$, $\gamma_{0}=2.4$. As shown in \citet{Dayal2015}, the fit well describes the scaling relation in CDM and WDM models (with $m_{\rm WDM}\geq 3 \kev$) towards $M_{\ast} \sim 10^{7}\msun$. It is also in good agreement with estimates using abundance matching \citep[e.g.][]{Kuhlen2012, Schultz2014} and direct observational estimates for more massive Lyman-break galaxies \citep[e.g.][]{Grazian2015,Song2016,Stefanon2017}. Empirically, we find $\gamma_{0}=1.9$ gives the best prediction on where the suppression on UV luminosity functions shows up. We use this relation to estimate the characteristic UV luminosity where differences between DM models show up.

\section{Estimating cosmic variance}
\label{app:cosmic-var}

In Section~\ref{sec:hmf-smf}, we describe the subsampling approach to estimate cosmic variances. Here we provide details on estimating the variance inflation factor. The halo mass function in a biased density field is known as the conditional halo mass function. As shown in e.g. \citet{Mo1996,Trapp2020}, the normalization of the conditional halo mass function at $(M,z)$ scales linearly (when $M$ is much smaller than the total mass in the volume) with the mass overdensity of the volume as $n(M,z)/\bar{n}(M,z) = 1 + b(M,z)\, \delta_{\rm m}$, where $b(M,z)$ is the bias function and $\bar{n}$ is the mean halo mass function. Therefore, the variance of the mass function $\log{n(M,z)}$ equals the variance of the overdensity field multiplied by $b(M,z)$.

The ratio between the total variance of the overdensity field over what we measure from subsampling 
\begin{align}
    I & = \sqrt{ \dfrac{\sigma^{2}(l=4\Mpc/h)}{\sigma^{2}(l=4\Mpc/h) - \sigma^{2}(l=8\Mpc/h)} } \nonumber \\
    & = \sqrt{ \dfrac{\int^{1/(4\Mpc/h)}_{0} \Delta(k)\,{\rm d}k/k}{\int^{1/(4\Mpc/h}_{1/(8\Mpc/h)} \Delta(k)\,{\rm d}k/k)} } \sim 1.3
\end{align}
The value does not vary in altDM models (since $4 \Mpc/h$ is much above the suppression scale). The variance of the halo mass function linearly scales with the variance of the overdensity field and therefore can be estimated as the variance calculated from subsampling times this inflation factor, $\sqrt{{\rm Var}_{\rm sample}[\log{n(M,z)}]} \times I$. 


\bsp	
\label{lastpage}
\end{document}